\documentclass[times]{speauth}

\usepackage{amssymb}
\usepackage{amsthm}

\usepackage[flushmargin, hang, bottom]{footmisc}

\usepackage{url}
\usepackage{algorithmic}
\usepackage[lined, linesnumbered]{algorithm2e}
\usepackage{amsmath} 
\usepackage{multirow}
\usepackage{subfig}
\usepackage{rotating}
\usepackage{listings}
\usepackage{float}
\usepackage{array}
\usepackage{epstopdf}
\usepackage{arydshln}

\theoremstyle{definition}

\theoremstyle{definition}
\newtheorem{MyDef}{Definition}

\usepackage{alltt}
\usepackage{textcomp}
\usepackage[T1]{fontenc}
\usepackage{float}
\usepackage{caption}

\begin{document}

\runningheads{L.~Iribarne \textit{et al.}}{Modeling Big Data based Systems through Ontological Trading}

\title{Modeling Big Data based Systems through Ontological Trading}

\author{Luis~Iribarne{\corrauth}, Jos\'e Andr\'es Asensio, Nicol\'as~Padilla, Javier Criado}

\address{Applied Computing Group, Department of Informatics, University of Almer\'ia (Spain)}

\corraddr{Luis Iribarne, Applied Computing Group, Department of Informatics, University of Almer\'ia, Spain\\ Email: luis.iribarne@ual.es}

\begin{abstract}
One of the great challenges the information society faces is dealing with the huge amount of information generated and handled daily on the Internet. Today, progress in Big Data proposals attempt to solve this problem, but there are certain limitations to information search and retrieval due basically to the large volumes handled, the heterogeneity of the information and its dispersion among a multitude of sources. 
{\color{black}In this article, a formal framework is defined to facilitate the design and development of an Environmental Management Information System which works with an heterogeneous and large amount of data. Nevertheless, this framework can be applied to other information systems that work with Big Data, since it does not depend on the type of data and can be utilized in other domains.}  
The framework is based on an Ontological Web-Trading Model (OntoTrader) which follows Model-Driven Engineering and Ontology-Driven Engineering guidelines to separate the system architecture from its implementation. The proposal is accompanied by a case study, SOLERES-KRS, an Environmental Knowledge Representation System designed and developed using Software Agents and Multi-Agent Systems. 
\end{abstract}


\maketitle

\section{Introduction}
\label{sec:introduction}

New Big Data techniques \cite{bayard} \cite{alexander} are starting to come into use to develop \textit{Information Systems}, and especially \textit{Web-based Systems}, for solving one of the major problems facing our information society, the huge amount of information managed in almost any area of society. Current information system development rules promote elimination of space-time barriers and facilitate asynchronous, ubiquitous communication with their wide variety of users (administrators, technicians, politicians, etc.), providing them with interactive information and graphical resources to assist them in problem-solving and decision-making. Even so, these users often experience information overloads, and must discriminate information in contents. This is because information search and retrieval must confront the huge and heterogeneous volume managed, and integrate many sources or the repositories where it is stored. Many techniques in different areas have to be applied to address this problem: information representation, information storage and retrieval, information filtering, studies on information search behavior, human-computer interaction, etc. 

Furthermore, current information system development requires integrity, scalability, accessibility, adaptability, extendability, etc., making rules and standards necessary for their construction. In particular, the design of Web-based Information Systems which handle large volumes of information, such as Environmental Management Information Systems (like the SOLERES system, used as an experimental example in this article), must use standard methods and techniques enabling a common vocabulary to represent the knowledge and real-time interaction mechanisms between user and system, and between system and third-parties or even components of the system itself. This has led to the use of new open distributed paradigms \cite{r59} and concepts such as Web semantics, data mining and query components related to this type of system \cite{r51}. One of their intrinsic characteristics is management of very disparate sources of knowledge for frequent use by a wide variety of users (administrators, technicians, politicians, etc.) who must cooperate with each other for better decision-making. 

This article presents a framework for design and development of information systems that rely on some of the main aspects of Big Data, i.e., the volume, velocity and variety of the information managed \cite{madden2012from}. Specifically this framework is focused on solving problems related to the heterogeneity (i.e., variety) of the data through the use of mediation mechanisms between exporters and importers of such information.
{\color{black} Nevertheless, volume and velocity issues are also addressed in this approach since the proposed infrastructure has been developed in order to manage a large amount of data and to perform query and retrieval operations in assumable times of Web domain (for example, balancing the workload on different nodes of the infrastructure).}
Previous conditions  result in a proposal for modeling systems which are based on Big Data concepts, although they may not be considered (from a strict point of view) as Big Data systems. The framework proposed, called Trading-based Knowledge Representation System (TKRS) is constructed following the rules and guidelines of Model-Driven Engineering (MDE) and Ontology-Driven Engineering (ODE), thereby enabling separation of the system architecture definition from its implementation. In TKRS, the system is considered (in the design stage) a distributed architecture consisting of a set of nodes (or ambients) which are in turn made up of a series of modules with certain preset functionalities for system deployment. A graphical editor implemented specifically for designing the architecture makes it possible to obtain a system model in a few simple steps. Node functionality is defined in the design stage as a type of capability the system must have, and may be implemented in different programming languages in the development stage. The TKRS framework has a system component implementation repository model connecting each component in the architecture to its corresponding implementation. TKRS uses model-to-model (M2M) transformations to relate architecture models to implementations in a system configuration model. Based on the configuration model, a concrete instance of the system (executable code) is acquired for the specific platform selected by applying a model-to-text (M2T) transformation. 

The keystone of the TKRS framework is its three-level data model. The framework considers the original data sources, which reside in their places of origin, first. The system manages part of the data from these original external data sources to process information related to its business logic. For example, in digital document data systems, the sources may reside in proprietary repositories in different entities or organizations. Each document is registered in the TKRS as a summary sheet (or template) of the document (or metadata). The metadata are modeled as a business logic information ontology. Therefore, each document that is entered in the system represents an instance of the ontology. As a search and retrieval mechanism, the TKRS uses a trading model called OntoTrader \cite{r31bis} which extends the traditional ODP \cite{r32} trading function. A trader (and by extension our OntoTrader) is a yellow-page service that handles a small collection of data for locating more detailed information. In TKRS, the OntoTrader trading model manages a meta-metadata repository of templates or documents (metadata). This trading service integrates the available information sources and system components, and improves information search and retrieval. Interaction with different trading model interfaces is defined by a series of service/process ontologies. The term ontological trading model comes from the use of these ontologies with those for data. 

In this article, as an experimental case study, the TKRS framework is applied in SOLERES, a Web-based Environmental Management Information System that basically manages cartography and satellite images. In this system, the business logic consists of processing the information to make cartographic information and satellite information correspond. This correspondence involves previously processing the cartography, dividing it into sectors, and classifying the satellite image to be matched. Cartography and its sectorization, and the satellite image and its classification are modeled as an ontology and stored in TKRS as an instance of this ontology. A data subset of this instance is registered in the trader to facilitate search, location and retrieval of information in the system. The SOLERES system emerged from a multidisciplinary research project funded by the Spanish Ministry of Science and Innovation and was supported by the application, integration and development of studies in satellite imaging, neural networks, cooperative systems based on multi-agent architectures, and software systems based on commercial components (third-party components). The goal of the project was to study the possibility and feasibility of correlating environmental variables used in creating ecological maps (based on cartography) with satellite or flight information, as was finally demonstrated. This correlation enables automatic creation of ecological maps in far less time and at lower economic and human costs than what has been done to date. For more information on this project see \url{http://acg.ual.es/soleres}. 

\begin{figure}[!b]
\begin{center}
\includegraphics[width=0.945\textwidth]{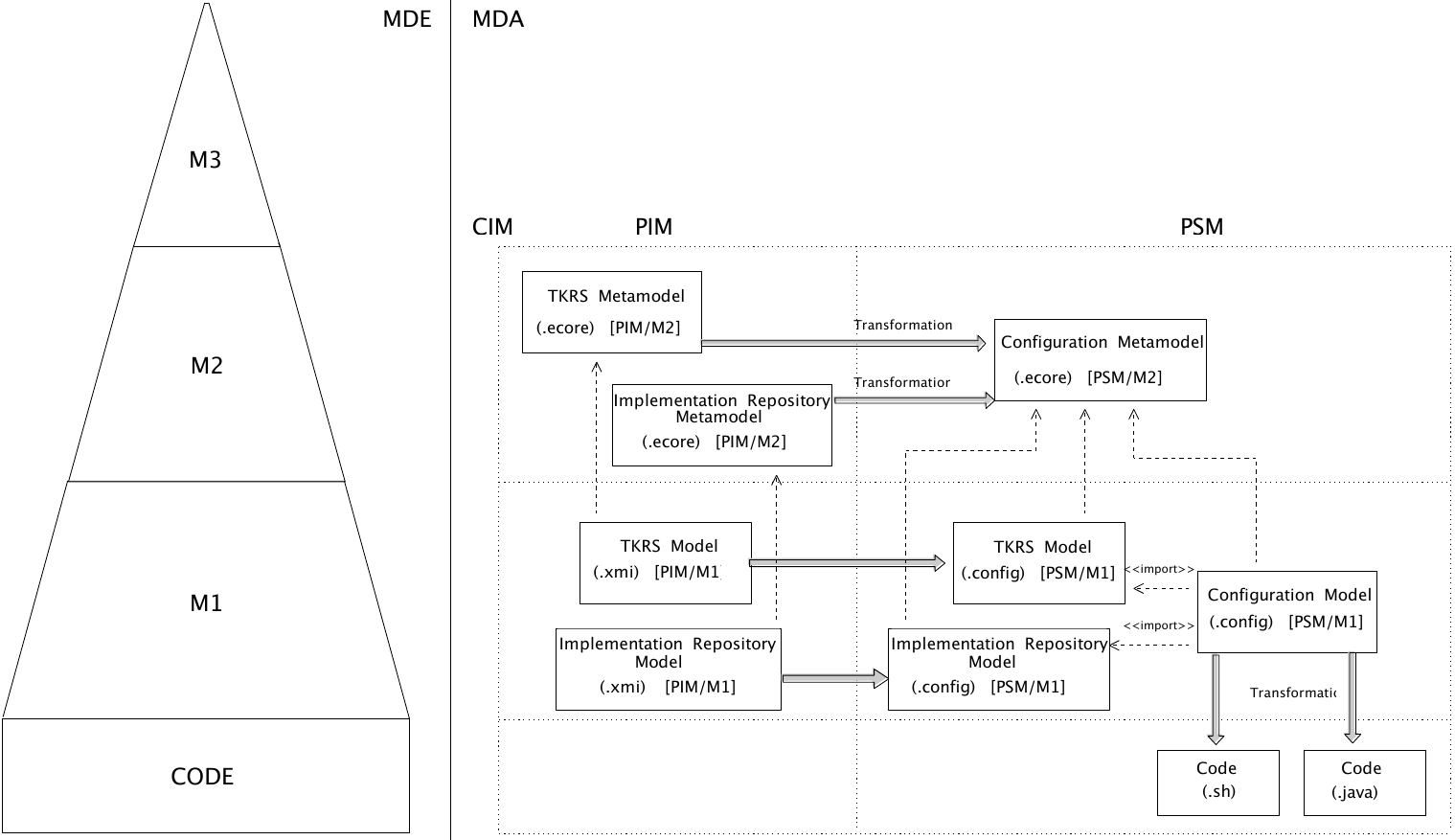}
\end{center}
\caption{Framework for design and development of TKRS systems}
\label{f1}
\end{figure}

As mentioned above, the TKRS framework was constructed following an MDE development methodology. Figure \ref{f1} shows a summary of its architecture. Three basic types of models are managed in it: (a) architecture models, (b) implementation repository models, and (c) configuration models. The first step is to make a \textit{Platform Independent Model} (PIM) of the system architecture (in M2 level, \textit{i.e.}, PIM/M2) and the implementation repository (also in PIM/M2) to be modeled by defining the two metamodels. Then, system configuration is modeled using a \textit{Platform Specific Model} (PSM) also in M2 level (PSM/M2), and it relates the architecture to a concrete implementation. For this, first a M2M transformation must be performed of both the architecture model (PIM/M1 to PSM/M1) and the repository model (also PIM/M1 to PSM/M1). Finally, based on the configuration model (PSM/M1), another transformation, now M2T, generates the implementation code for the specific platform. Eclipse-based technologies, such as \textit{Eclipse Modeling Framework} (EMF), \textit{Graphical Modeling Framework} (GMF), \textit{XText} and \textit{Xpand} are used to develop it. Thus, the metamodels are defined by a graphical editor and stored in ``.ecore'' files, the models created based on them are represented in ``.xmi'', and the configuration models are expressed by a Domain Specific Language (DSL) which generates ``.config'' files. 

The rest of the article is organized as follows. The three main parts of the framework, (a) the architecture, (b) the implementation repository, and (c) the configurations, are modeled in Section \ref{sec:framework}. Then, a mathematical formalization is presented in Section \ref{sec:formalization}. In Section \ref{sec:casestudy}, the SOLERES-KRS case study, in which the TKRS proposal is validated, is presented. Section \ref{sec:rw} describes some related studies. Finally, in Section \ref{sec:conclus}, some conclusions and future work are discussed.

\section{Modeling the TKRS framework}
\label{sec:framework}

As the TKRS framework uses an MDE approach with model transformation, several metamodels had to be designed for the framework proposed. A transformer is a process that uses an M1 input model according to a preset MM1 metamodel, and transforms it into M2 according to its MM2 metamodel based on a set of preset rules. As mentioned above, the TKRS framework uses three basic models: architecture, implementation and configuration. It was therefore necessary to create a metamodel for each one, as described in the following sections. 

\subsection{Modeling the TKRS architecture}

\begin{figure}[!b]
\begin{center}
\includegraphics[width=9.5cm]{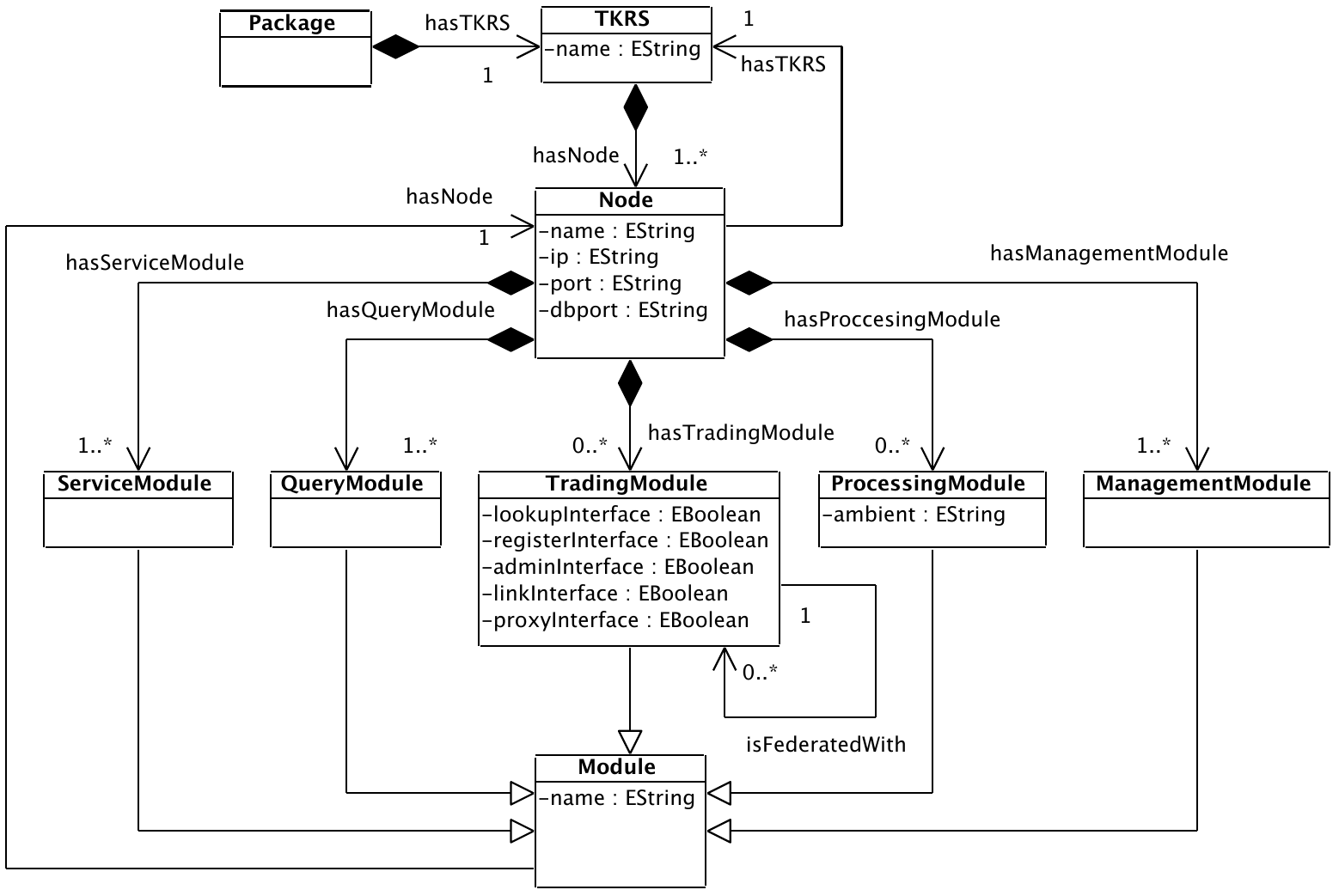}
\end{center}
\caption{TKRS architecture metamodel}
\label{f2}
\end{figure}

This section presents the TKRS architecture metamodel shown in Figure \ref{f2}. It may be observed that the TKRS is a type of distributed node architecture, where each node consists of a set of Modules, as described below. Each node has at least one service module with a set of services for registering modules and components, verifying states, etc. Each node also has a management module which mediates between the user interface and the rest of the system modules, managing user demands, and a query module related to solution of the user information queries. In addition, a node can also have one or more trading modules, which enable the search and location of information in the system (and here is where the OntoTrader model comes in) although it is not an indispensable requirement, and one or more processing modules, responsible for the management of knowledge bases. These may be grouped in larger logical units called ambients by any type of criterion, such as location. The system must have at least one trading module and one processing module in one of its nodes. 

The metamodel proposed in Figure \ref{f2} (PIM/M2 in Figure \ref{f1}) was defined using EMF \cite{r22}, and TKRS models reflecting the concepts, relationships and restrictions can be created based on it. For example, the Node metaclass has an identifier (\textit{name}), an IP address (\textit{ip}) and two communication ports (\textit{port} and \textit{dbport}). It can also give the definition of an abstract Module metaclass, with an attribute \textit{name}, which the different types of modules inherit. The trading module, which provides the ontological trading service with form, sets five Boolean attributes that show whether it implements the various interfaces or not: \textit{Lookup}, \textit{Register}, \textit{Admin}, \textit{Link} and \textit{Proxy}. The Lookup and Register interfaces are necessary, because each trading module is linked to a metadata repository. This metamodel also defines trading module federation by the \textit{isFederatedWith} relationship, and a connection between each processing module and each query module, by means of the \textit{usesRegisterInterface} and \textit{usesLookupInterface} relationships, respectively, with a trading module. These relationships are analyzed in Subsection \ref{s31}.

The creation of TKRS models (PIM/M1 in Figure \ref{f1}) using the metamodel described, is done by a GMF graphical model editor, which along with OCL (\textit{Object Constraint Language}) rules, enables all the restrictions to be represented. For a more detailed description of this tool see \cite{r5}. 

\subsubsection{Integrating the ontological trading service.}
\label{s31}

\begin{figure}[!b]
\begin{center}
\includegraphics[width=0.86\textwidth]{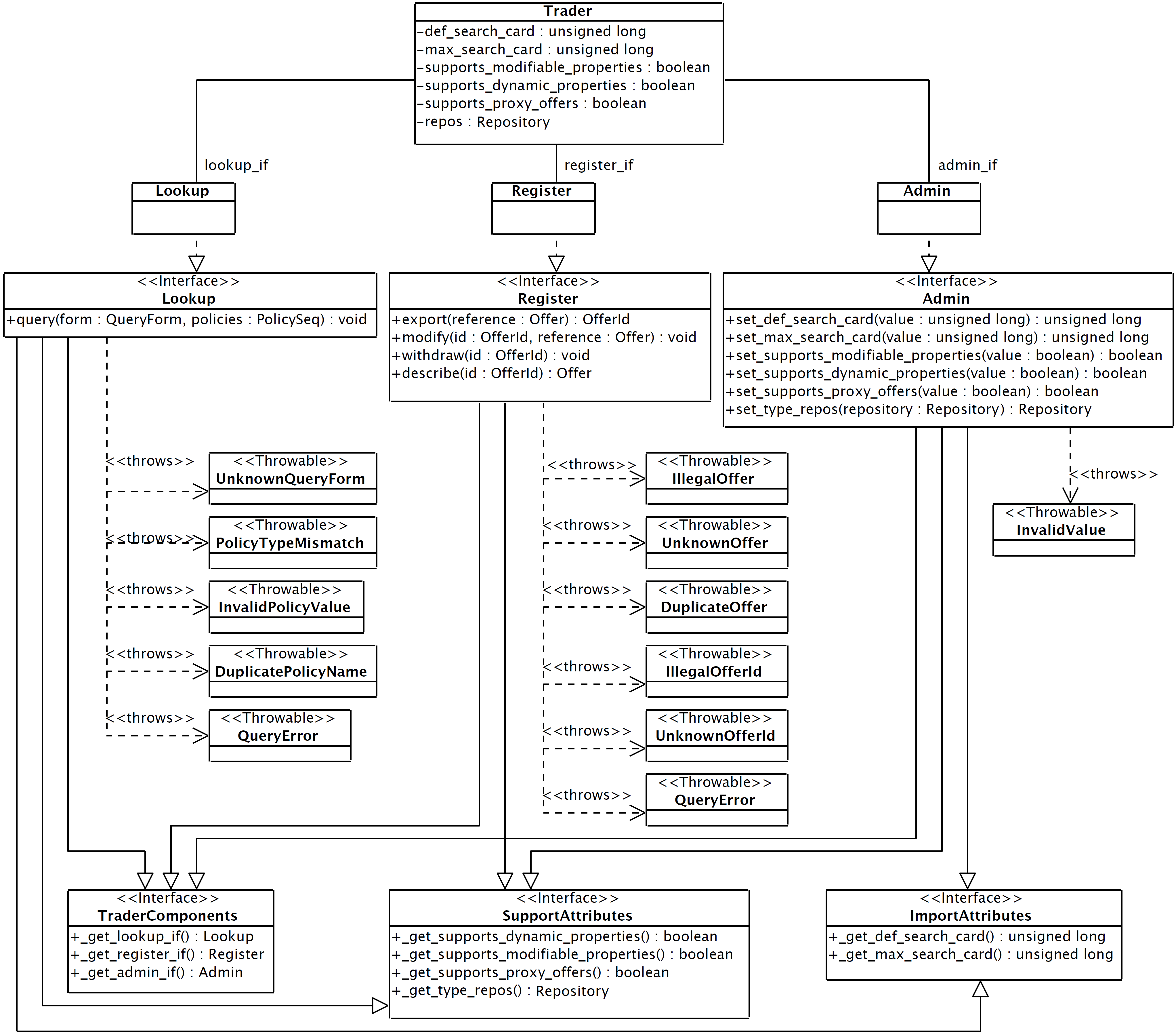}
\end{center}
\caption{Conceptual model of stand-alone trading}
\label{f4}
\end{figure}

One of the characteristics of TKRS design is that the OntoTrader model is the axis of its architecture. As already mentioned, the ontological trading model proposed is based on the OMG ODP specification \cite{r32} and has been adapted to handle documents with metadata of any type of information not just service specifications. This model is a stand-alone trading service that implements the Lookup, Register and Admin interfaces, as shown in the conceptual model shown in Figure \ref{f4}. The Lookup interface is only concerned with information queries, the Register interface enables full management of the service information, and finally, configuration parameters and trading policies can be modified from the Admin interface. 

The service/process ontologies are used to define the behavior and interaction protocols of objects or components interacting with the trading service interfaces. All the ontologies were modeled as concepts (domain entities), actions performed in the domain (actions which affect concepts) and predicates (expressions related to concepts), represented in UML class diagrams by the stereotypes $<<$Concept$>>$, $<<$Action$>>$ and $<<$Predicate$>>$, respectively. Each of these is analyzed in more detail in the following Sections. 

The ontological trading service is encapsulated in the TKRS architecture trading module, and the service/process ontologies (Lookup, Register and Admin) use the rest of the modules to interact with it, and even to communicate with each other. Table \ref{t2} shows the ontologies that each of the TKRS system modules use. The influence of the OntoTrader model on management and information query is analyzed below.

\begin{table}
\centering
\caption{Use of service/process ontologies in TKRS}
\label{t2}
{\fontsize{8pt}{9pt}\selectfont
\begin{tabular}{llp{4cm}}
\hline
Source & Ontology & Target\\
\hline
Management module & Lookup Ontology & Query module\\
Query module & Lookup Ontology & Trading module\\
Query module & Lookup Ontology & Processing module\\
Query module & Lookup Ontology & Management module\\
Trading module & Lookup Ontology & Query module\\
Processing module & Lookup Ontology & Query module\\
\hline
Management module & Register Ontology & Processing module\\
Processing module & Register Ontology & Trading module\\
\hline
Management module  & Admin Ontology & Trading module\\
\hline
\end{tabular}
}
\end{table}

Four actions are implicit in information management: data insertion, modification, elimination and description. All of them originate from a query made from the interface by a user or a group of users. This query is picked up by the management module and transferred to the processing module specified by the user: (a) if data are to be inserted, this module stores the metadata in a local repository and generates a new meta-metadata record from them, which stores it in a local meta-metadata repository, and also transfers them to a trading module with which it is associated, concluding the action. Thus, the trading module has an overall repository with all the meta-metadata of all the processing modules associated with it (fundamental to information search and retrieval); (b) if data are to be modified or deleted, first the local processing module repositories are affected, and then the action is transferred to the trading module so it can act accordingly, the same as for insertion; (c) data description is solved directly in the processing module, which returns the information demanded by the user based on which it is stored in its metadata repositories. 

\begin{figure}
\begin{center}
\includegraphics[width=6.5cm]{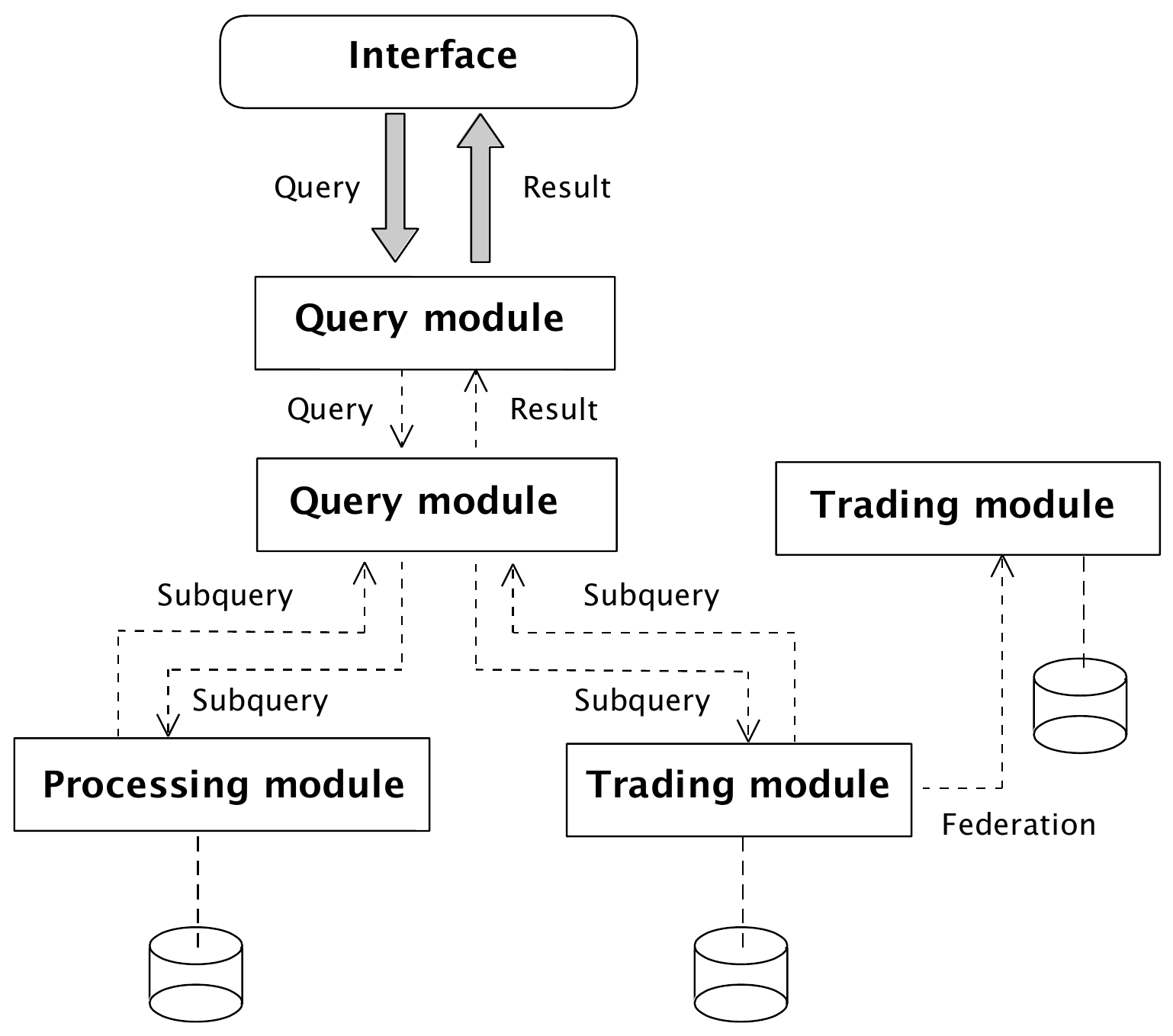}
\end{center}
\caption{TKRS operating diagram during a query}
\label{f5}
\end{figure}

The OntoTrader trading service information retrieval process follows a ``Query-Search/Recovery-Response'' mechanism (QS/RR), based on a three-level client/server model $<O, M, I>$. This model is configured in the TKRS as follows (summarized in Figure \ref{fig:qsrr}): on Level 1 (N1) are queries generated and processed by a component or object in the user interface and received in the system by management modules O, on Level 3 (N3) are the processing modules, and their metadata repositories I, and middleware M, which enables location of information sources, are on Level 2 (N2), where the query and trading modules with their meta-metadata repositories operate. As an indispensable premise for it to function, each object must be associated with a trading service, which means that each query module and each processing module must be associated with a trading module, and is the reason why each processing module and each query module in the metamodel defined in Section \ref{sec:framework} is connected to a trading module (by means of the \textit{usesRegisterInterface} and \textit{usesLookupInterface} relationships, respectively). The management modules are mere organizers transmitting queries to the query modules. The reflection, delegation and federation scenarios, summarized graphically in Figure \ref{f5}, would be: 

\begin{figure}[!b]
\begin{center}
\includegraphics[width=9.585cm]{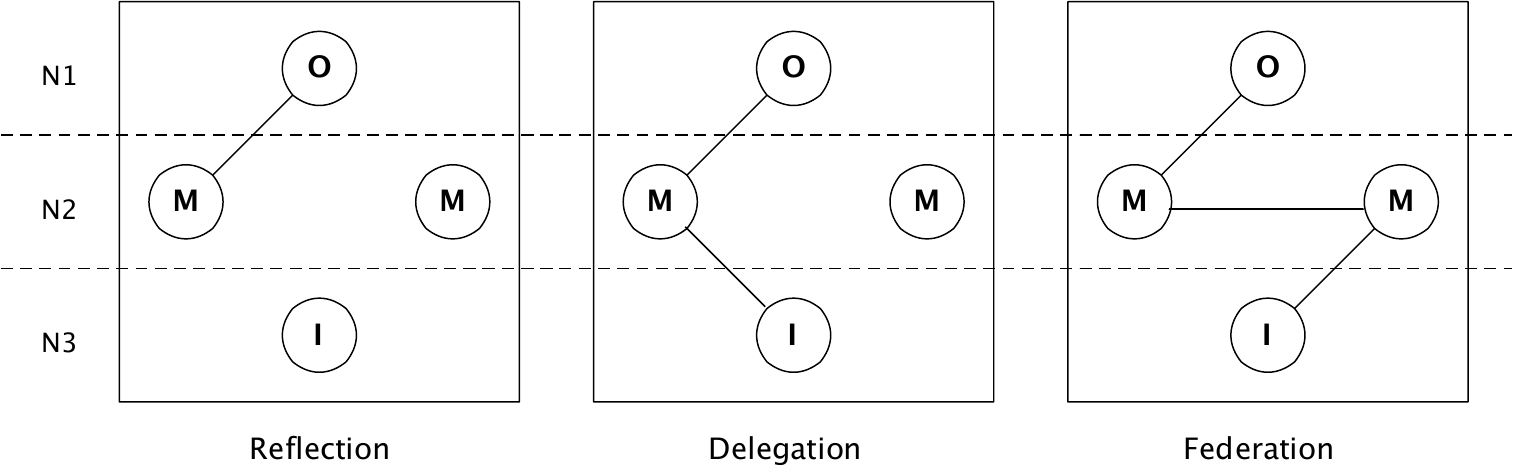}
\end{center}
\caption{OntoTrader ontological trading service operation models}
\label{fig:qsrr}
\end{figure}

\begin{itemize}
\item 
In the reflection model, the query generated in interface O can be satisfied with meta-metadata documents stored in the repository associated with a trading module M. The query would then be transferred from the interface to a management module, and from there to a query module, which is the one that really plans it, and then to the trading module. In this case, the model's $<O, M>$ pair intervenes. 

\item 
The delegation model negotiates with the trading service indirectly. The query generated in interface O goes to a query module that breaks it down into two or more subqueries. One transfers it to the associated trading module M to locate the origin of the information in its repository, and the query module reformulates the subquery or subqueries and transfers it to the processing module repositories I indicated by the other. In this case, elements from $<O, M, I>$ intervene.

\item 
In the federation model, two or more trading modules are configured to use a delegation mechanism. As in the first case, the query from interface O arrives at the first trading module associated with query module M it comes across, which propagates it to another trading module M with which it is federated and will locate the external information sources I. In this case, the intervening elements are $<O, M, M, I>$. 
\end{itemize}

\subsubsection{Trading ontologies.}

The three main ontologies used for trading are Lookup, Register and Admin, as defined below. 

\paragraph{\bf{Lookup Ontology:}} 

\begin{figure}
	\begin{center}
		\includegraphics[width=1.00\textwidth]{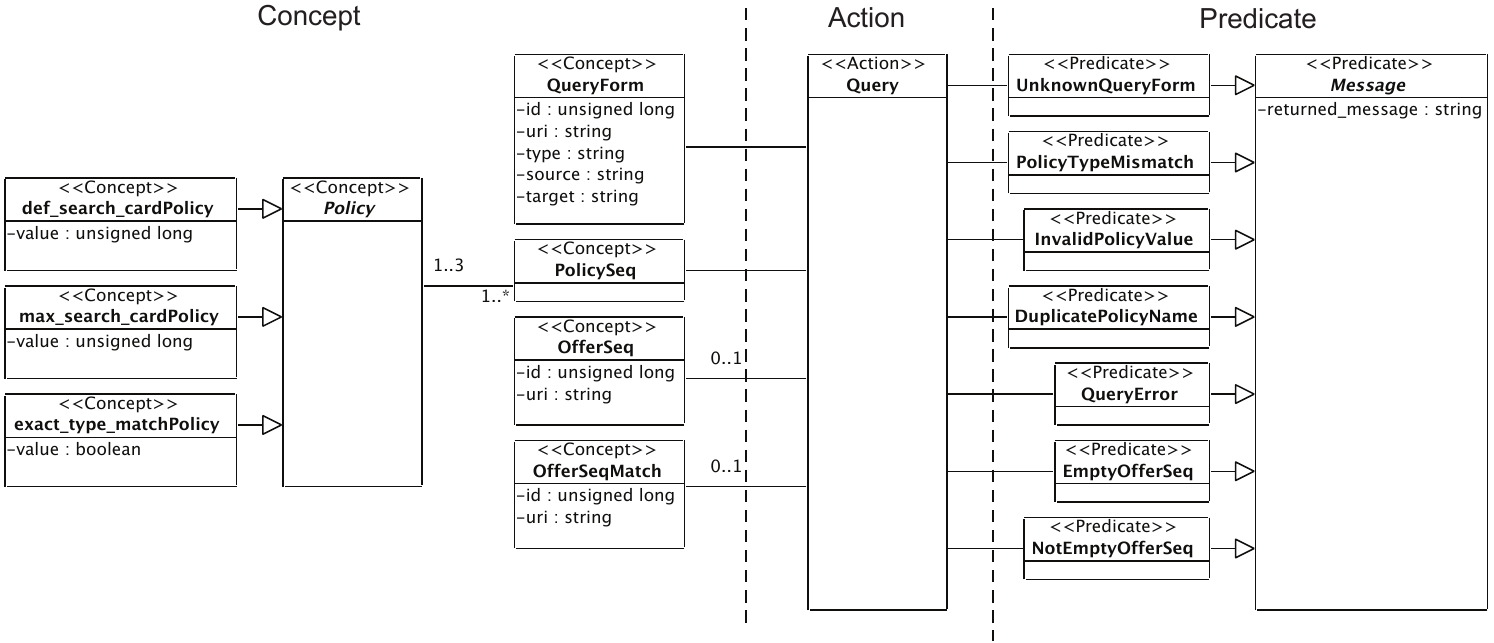}
	\end{center}
	\caption{{\color{black}Lookup ontology conceptual model}}
	\label{f6}
\end{figure}

The Lookup ontology searches for and retrieves information in a specific repository following previously set criteria or policies. Figure \ref{f6} shows the structure of this ontology. The Query action makes use of the QueryForm and PolicySeq concepts. The first represents a query expressed in a specific language and has several properties: \textit{id}, \textit{uri}, \textit{type}, \textit{source} and \textit{target}. The property \textit{id} is a query identifier, \textit{uri} is a reference to the file that stores the query expressed in a certain language. Table \ref{t3} shows an example of a query in SPARQL \cite{quilitz2008sparql} in the EID repository in the SOLERES case study analyzed in Section \ref{sec:casestudy} to get ``the name of the variables used in the classifications made during 2008''. {\color{black} It is important to note that this query example is related to the type of information that is managed in the case study and, if different information is used, this code snippet must be changed accordingly.} The property \textit{type} shows the exact type of query (for example, if it is a query about the repository or another previous query). And \textit{source} identifies the object or component the query comes from, and finally, \textit{target}, enables its destination object or component to be identified. 

The PolicySeq concept also represents a set of query policies. The Policy is an abstract concept used to represent a specific policy by means of the tuple (\textit{name}, \textit{value}). Therefore, its only property is a \textit{value} representing a given Policy. The three policies defined are: \textit{def\_search\_cardPolicy}, \textit{max\_search\_cardPolicy} and \textit{exact\_type\_matchPolicy}. The first shows the number of documents to be located in the query, \textit{max\_search\_cardPolicy} and \textit{exact\_type\_matchPolicy} state whether the documents to be located must coincide exactly with the query conditions or not (true if the documents returned must meet the conditions set by the query 100\%, false if the documents located do not necessarily have to coincide exactly with the conditions). 

\begin{table}
\centering
\caption{Example query expressed in SPARQL}
\label{t3}
{\fontsize{6pt}{7pt}\selectfont
\begin{tabular}{p{10cm}}
\hline
\begin{verbatim}
 1  PREFIX eid: <http://www.ual.es/acg/ont/TKRS/EIDOntology.owl#>
 2  PREFIX rdf: <http://www.w3.org/1999/02/22-rdf-syntax-ns#>
 3  SELECT DISTINCT ?EID__VariableName
 4  WHERE
 5  {
 6    ?EID__EID eid:EID_eimId ?EID__EID_eimID .
 7    ?EID__EID eid:EID_hasClassification ?EID__Classification .
 8    ?EID__Classification eid:Classification_hasLayer ?EID__Layer .
 9    ?EID__Layer eid:Layer_hasVariable ?EID__Variable .
10    ?EID__Variable eid:Variable_name ?EID__VariableName .
11    ?EID__Classification eid:Classification_hasTime ?EID__Time .
12    ?EID__Time eid:Time_year ?EID__TimeYear .
13    FILTER (?EID__TimeYear = 2008) .
14  }
15  ORDER BY ASC(?EID__VariableName)
\end{verbatim}
\\
\hline
\end{tabular}
}
\end{table}

The output of the Query action may be a controlled exception or the result of the query itself, all of them represented by the abstract predicate Message they inherit from. This predicate has the property returned\_message, which contains the text of the output message. Exceptions that could arise are: \textit{UnknownQueryForm}, \textit{PolicyTypeMismatch}, \textit{InvalidPolicyValue}, \textit{DuplicatePolicyName} and \textit{QueryError}. The exception \textit{UnknownQueryForm} shows that the query could not be answered because the file specified in the uri was not accessible; \textit{PolicyTypeMismatch} shows that the type of Policy value was not right; \textit{InvalidPolicyValue} shows that the Policy value was not within the permissible range for this Policy; \textit{DuplicatePolicyName} shows that more than one value for the same Policy within PolicySeq was specified; and finally, the exception \textit{QueryError} shows that there was an error in executing the query. 

When the query is made successfully, the predicate \textit{EmptyOfferSeq} is employed if no document is returned, or \textit{NotEmptyOfferSeq} if it is. This predicate in turn makes use of the OfferSeq and OfferSeqMatch concepts to store the set of documents located and the percentage of coincidence of each document found, respectively, depending on the conditions of the query. The OfferSeq concept has the properties of the query \textit{id} and the \textit{uri} of the file that stores the documents found, just as the OfferSeqMatch concept, where the uri now refers to another file which contains a set of documents in the form of the tuples $<$\textit{identifier\_of\_record\_found}, \textit{percentage\_of\_coincidence}$>$. 

Table \ref{t4} summarizes the concepts, actions and predicates in the Lookup ontology. Abstract predicates Message and Policy are not included. 

\begin{table}
\centering
\caption{Lookup, Register and Admin ontology elements}
\label{t4}
{\fontsize{7pt}{7pt}\selectfont
\begin{tabular}{|l|l|l|l|}
\hline
{\bf Ontology} & {\bf Concepts} & {\bf Actions} & {\bf Predicates}\\
\hline
Lookup & QueryForm & Query & UnknownQueryForm\\
& PolicySeq & & PolicyTypeMismatch\\
& def\_search\_cardPolicy & & InvalidPolicyValue\\
& max\_search\_cardPolicy & & DuplicatePolicyName\\
& exact\_type\_matchPolicy & & QueryError\\
& OfferSeq & & EmptyOfferSeq\\
& OfferSeqMatch & & NotEmptyOfferSeq\\
\hline
Register & Offer & Export & IllegalOffer\\
& OfferId & Modify & UnknownOffer\\
& & Withdraw & DuplicateOffer\\
& & Describe & IllegalOfferId\\
& & & UnknownOfferId\\
& & & QueryError\\
& & & ExportedOffer\\
& & & ModifiedOffer\\
& & & WithdrawnOffer\\
& & & NotDescribedOffer\\
& & & DescribedOffer\\
\hline
Admin & Def\_search\_card & SetDef\_search\_card & ModifiedDef\_search\_card\\
& Max\_search\_card & SetMax\_search\_card & ModifiedMax\_search\_card\\
& Offer\_repos & SetOffer\_repos & ModifiedOffer\_repos\\
& & GetDef\_search\_card & ReturnedDef\_search\_card\\
& & GetMax\_search\_card & ReturnedMax\_search\_card\\
& & GetOffer\_repos & ReturnedOffer\_repos\\
& & & InvalidValue\\
\hline
\end{tabular}
}
\end{table}

\paragraph{\bf{Register Ontology:}}
\label{s312}

\begin{figure}[!b]
\begin{center}
\includegraphics[width=0.717\textwidth]{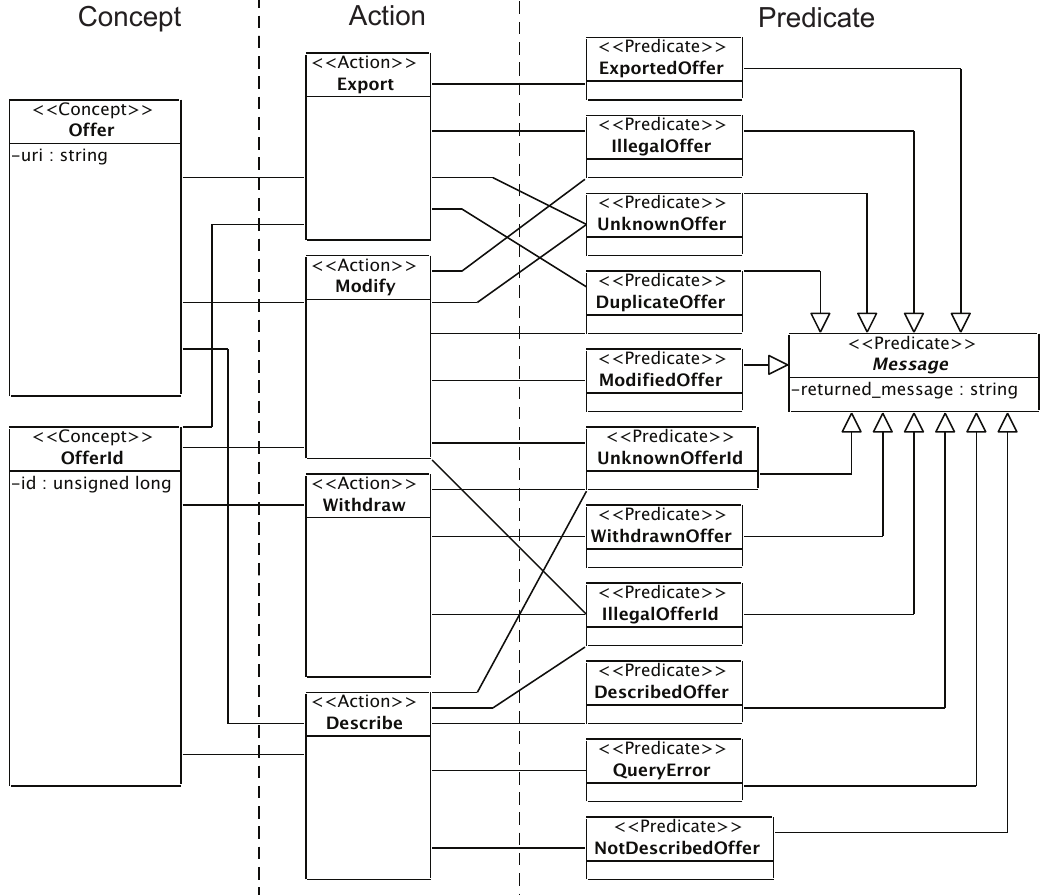}
\end{center}
\caption{{\color{black}Register ontology conceptual model}}
\label{f7}
\end{figure}

The metadata documents stored by the trading service in its repository are managed by the Register ontology which supports the creation, modification, elimination and description of documents. This ontology includes the Export, Modify, Withdraw and Describe actions. The Export action inserts a document previously specified in the Offer concept, which is stored in a file referenced by the concept's \textit{uri} property. The Modify action represents a document modification of a document and uses the Offer concept to refer to the new document with the changed information and the OfferId concept to represent the identifier of the document that is going to be modified. Finally, the Withdraw and Describe actions represent the elimination and description, respectively, of a particular document specified by the OfferId concept. Figure \ref{f7} shows the structure of the Register ontology. 

The result of an action may produce a controlled exception in the system or a valid result. Both are represented by a predicate that inherits from the abstract predicate Message, as in the Lookup ontology. The exceptions that can occur are \textit{IllegalOffer}, \textit{UnknownOffer}, \textit{DuplicateOffer}, \textit{IllegalOfferId}, \textit{UnknownOfferId} and \textit{QueryError}. The first three can come from the actions Export and Modify: the exception \textit{IllegalOffer} shows that the metadata document that it is intended to insert or modify is incorrect, \textit{UnknownOffer} shows that the document cannot be found for its insertion or modification because the file specified in the \textit{uri} is inaccessible, and the exception \textit{DuplicateOffer} shows that the document already exists in the repository. The actions Modify, Withdraw and Describe can also generate the exceptions \textit{IllegalOfferId} and \textit{UnknownOfferId}: The first shows that the \textit{id} of the document that it is intended to modify, eliminate or describe is incorrect, while the second shows that there is no document in the repository with this id for its modification, elimination or description. Finally, the Describe action can generate the exception \textit{QueryError}, which shows that there has been an error in making the query for the document specified for its description. 

The ontology also defines the \textit{ExportedOffer}, \textit{ModifiedOffer} and \textit{WithdrawnOffer} predicates to show that the Export, Modify and Withdraw actions, respectively, have been performed successfully. The \textit{ExportedOffer} predicate makes use of the OfferId concept to show the id assigned to the new document. The \textit{NotDescribedOffer} predicate of the Describe action shows the document was not located, and \textit{DescribedOffer} that it was. This predicate makes use of the Offer concept to return the document requested. 

Table \ref{t4} also summarizes Register ontology concepts, actions and predicates. The abstract predicate Message is not included. 

\paragraph{\bf{Admin Ontology:}}
\label{s313}

\begin{figure}[!b]
	\begin{center}
		\includegraphics[width=13cm]{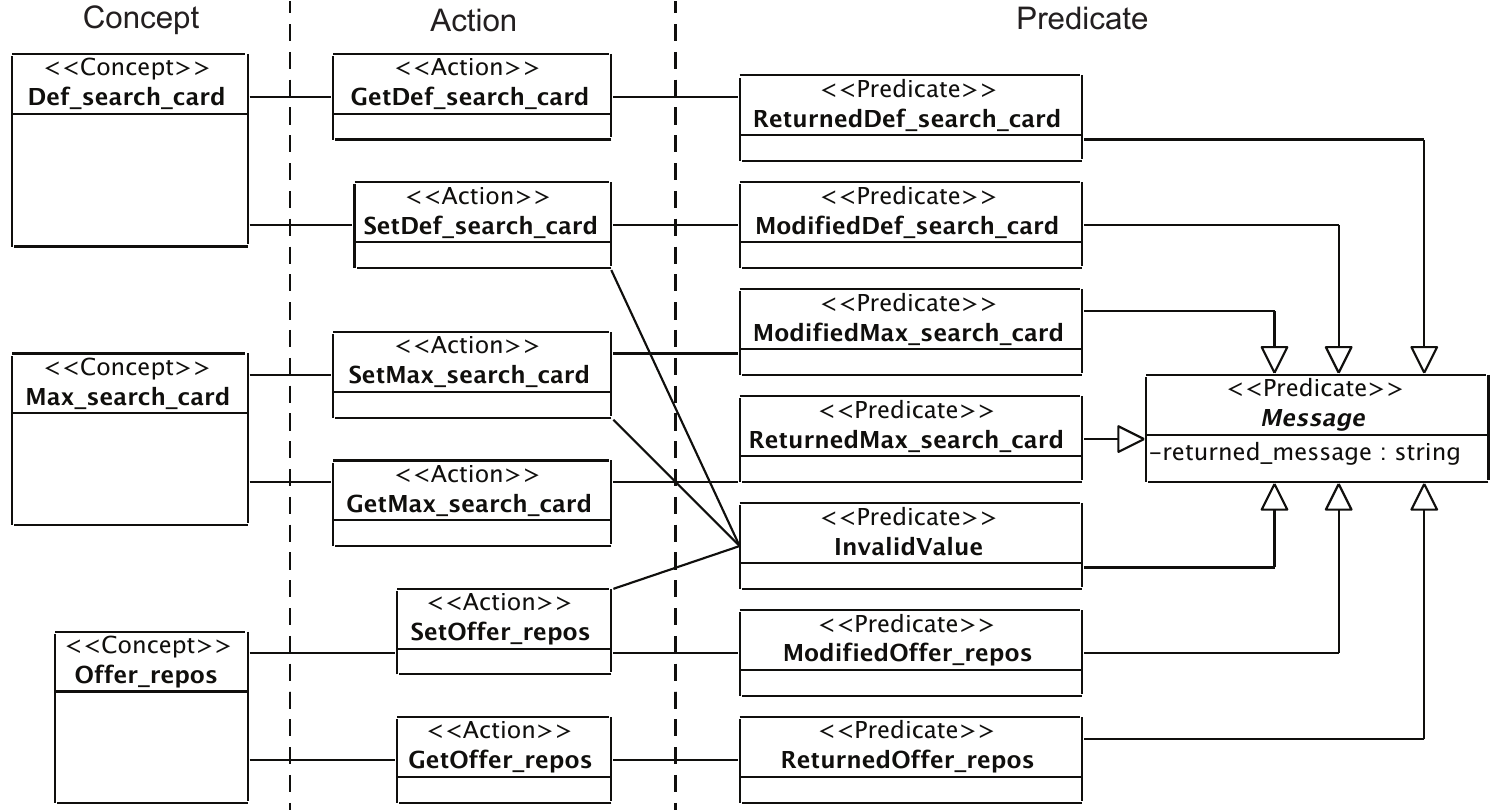}
	\end{center}
	\caption{{\color{black}Admin ontology conceptual model}}
	\label{f8}
\end{figure}

The Admin ontology is used for actions that modify the main configuration parameters and trading service policies. This ontology enables the maximum number of results found after executing a query to be modified, for example. Figure \ref{f8} shows its structure. This ontology defines three pairs of actions: \textit{SetDef\_search\_card}/\textit{GetDef\_search\_card} sets and obtains the predetermined number of metadata documents that are retrieved in a search and uses the \textit{Def\_search\_card} concept to store this information, \textit{SetMax\_search\_card}/\textit{GetMax\_search\_card} sets and obtains the maximum number of documents that are retrieved in a search and uses the \textit{Max\_search\_card} concept to store this information, and \textit{SetOffer\_repos}/\textit{GetOffer\_repos} can set/obtain the address (in URI format) of the repository with the metadata documents that uses the trading service (the \textit{Offer\_repos} concept is used to store the address).

When the ``Set'' actions are performed successfully, the trader returns the \textit{ModifiedDef\_search\_card} and \textit{ModifiedOffer\_repos} predicates to show that the values of the \textit{Def\_search\_card}, \textit{ModifiedMax\_search\_card} and \textit{Offer\_repos} parameters, respectively, have been set satisfactorily. On the other hand, it could return the \textit{ReturnedDef\_search\_card}, \textit{ReturnedMax\_search\_card} and \textit{ReturnedOffer\_repos} predicates to show that the values of the ``Get'' action parameters have been returned correctly. The only controlled exception that could occur is the InvalidValue predicate which could come from the actions \textit{SetDef\_search\_crad}, \textit{SetMax\_search\_card} or \textit{SetOffer\_repos}. This exception shows that the value specified for the parameter in question is invalid. As in the Lookup and Register ontologies, all the predicates inherit from the abstract predicate Message, which contains the property \textit{returned\_message} to store the output message text. 

Table \ref{t4} summarizes the concepts, actions and predicates in the Admin ontology. The abstract predicate Message is not included.

\subsubsection{Representing knowledge.}
\label{s32}

The TKRS can manage a large volume of data originating in one or more external sources directly related to one or more processing modules. These data originally make up the first level of the system data architecture. Although one of the main purposes of TKRS is to improve information search and retrieval, it is obviously inappropriate to work directly with the data given their probable volume, so a second level of metadata is generated and stored in the processing modules. The volume of these metadata is already much smaller, and although it may sometimes be necessary to recur to the lower level of data to solve a query, most of them can be solved on this higher level. However, should it be necessary, the metadata would act as a filter reducing the volume of information and therefore query execution time. 

\begin{figure}[!b]
\begin{center}
\includegraphics[width=11cm]{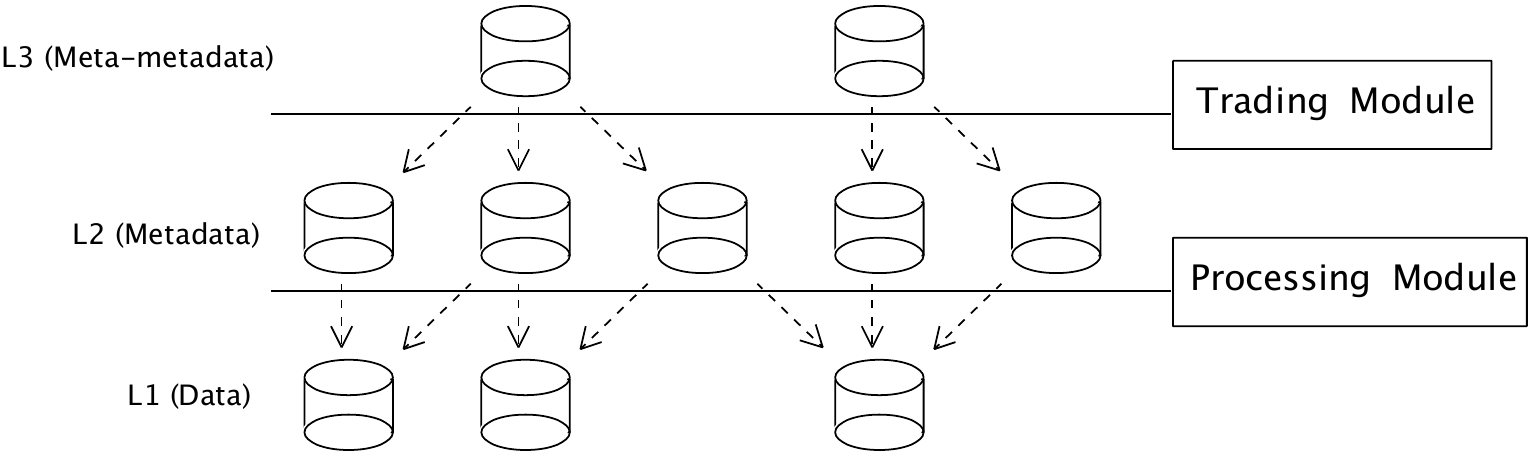}
\end{center}
\caption{TKRS three-level data architecture}
\label{f9}
\end{figure}

Since the system is distributed, a new level, called the meta-metadata level (which has certain information relating the different processing modules and is used by the trading modules), is included in the architecture. The ontologies have a basic role in the definition of these metadata and meta-metadata, in addition to the one they already have in defining system component interaction and communication protocols. Figure \ref{f9} shows the three levels of data architecture. 

\subsection{Modeling the TKRS implementation repository}
\label{s4}

\begin{figure}
\begin{center}
\includegraphics[width=4.5cm]{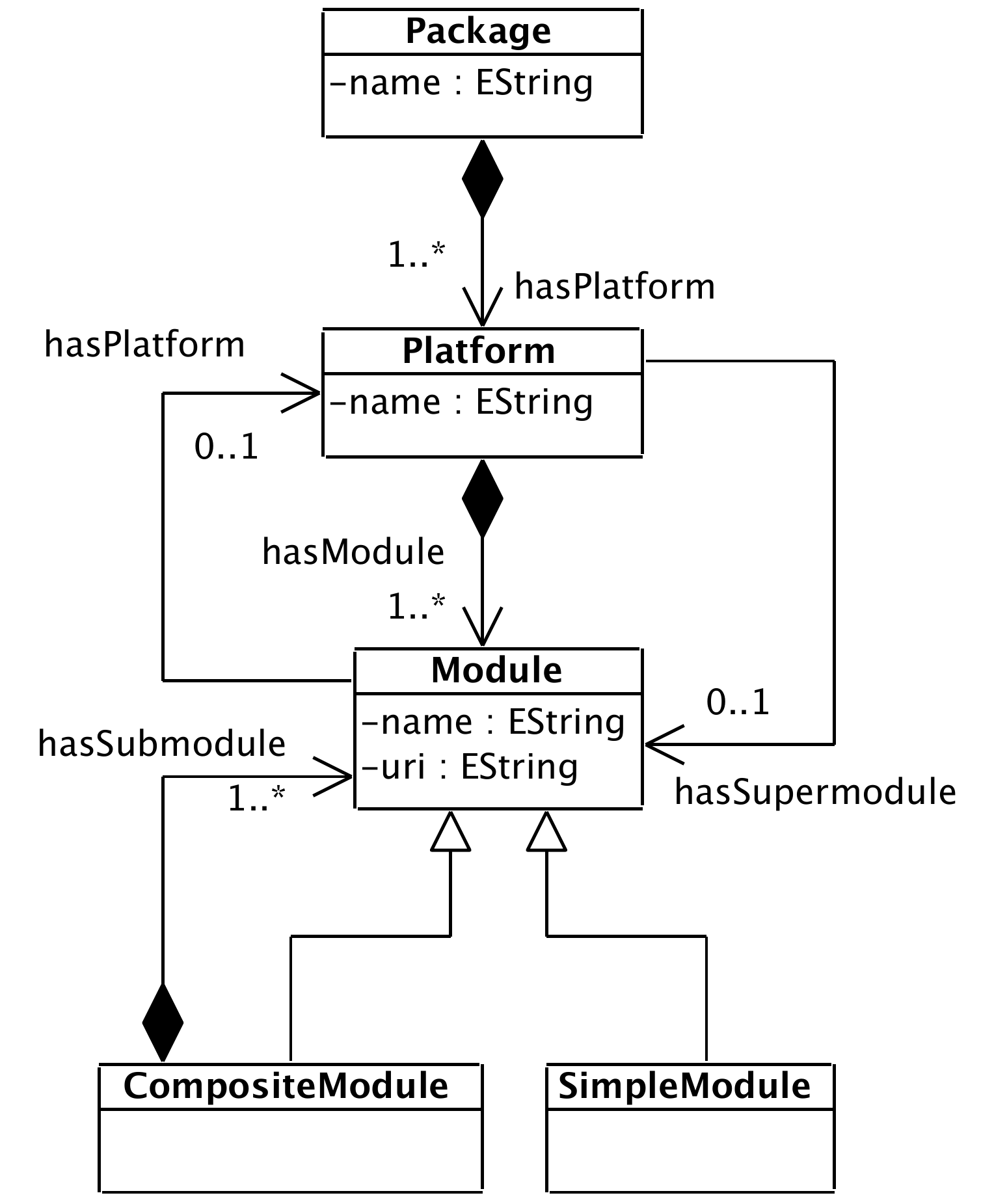}
\end{center}
\caption{Metamodel of a TKRS implementation repository}
\label{f10}
\end{figure}

After the TKRS architecture has been modeled (Metamodel PIM/M2 in Figure \ref{f1}), the next step is to model system implementation, since there may be more than one implementation depending on the development platform used, so they are all grouped together in a single repository. The structure of this implementation repository is very basic, and as observed in Figure \ref{f10}, which shows the repository metamodel (PIM/M2), it is made up of platforms (metaclass Platform) and implementations of each of the system modules in these platforms. The Platform metaclass includes a required attribute (name) specifying the name of the development platform used. A module (CompositeModule metaclass) can in turn be made up of one or more modules (CompositeModule or SimpleModule). The abstract metaclass Module (from which both simple and composite modules inherit) is also included, and two obligatory attributes, \textit{name} and \textit{uri}, specify the name of the module and a reference to the location of the implementation files, respectively. Finally, each module has a reference to its container module, if any, and its platform. This metamodel was again defined using EMF, and the reflective ecore model editor can be used to create concrete implementation repository models, implemented as analyzed in Section \ref{s62} based on the Multi-Agent Systems technology.

\subsection{Modeling the TKRS configuration}
\label{sec:modeling}

\begin{figure}[!b]
\begin{center}
\includegraphics[width=0.5\textwidth]{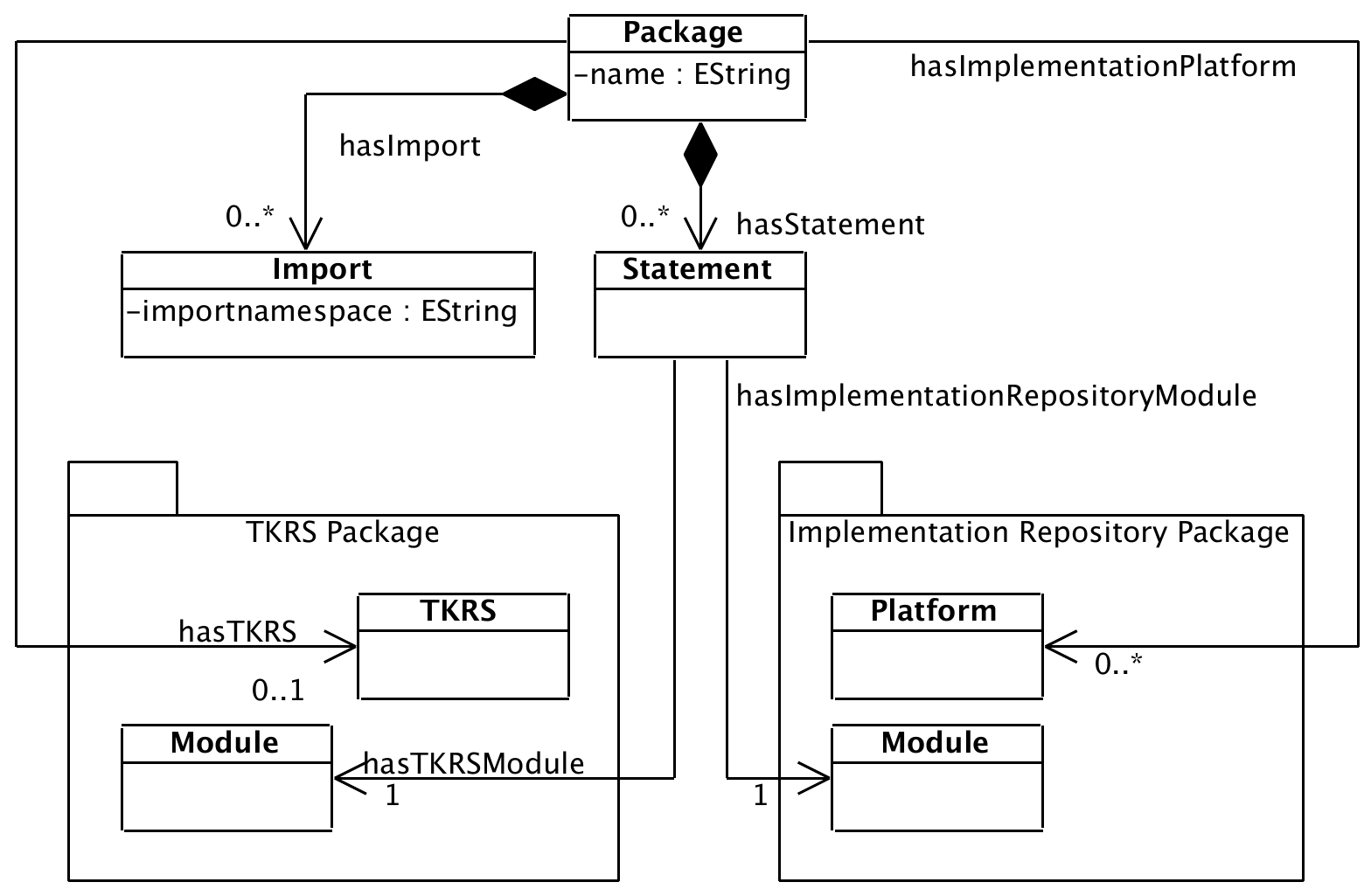}
\end{center}
\caption{TKRS configuration metamodel}
\label{f11}
\end{figure}

When the TKRS architecture and a repository of implementations on different platforms (Metamodel PIM/M2 in Figure \ref{f1}) have been defined, the next step for final system deployment in a given environment consists of setting up a mechanism to relate each element in the architecture to its implementation. This is exactly what is proposed below. As shown in Figure \ref{f11}, a new metamodel (PSM/M2) was defined in which the Statement metaclass relates each metamodel module in the TKRS architecture to the Module abstract metaclass in the implementation repository metamodel by means of the Module abstract metaclass. Furthermore, to be able to work with both the architecture and repository models, it includes the Import metaclass with the attribute \textit{importedNamespace} so the configuration models can import them. This metamodel, like the metamodels above, was defined using EMF, but to facilitate the creation of configuration models, a \textit{Domain Specific Language} (DSL) was developed using Eclipse XText. This tool can define a grammar based on a metamodel which describes both the concrete syntax of the language desired to implement and the procedure the parser must follow to generate a model. 

\begin{table}
\centering
\caption{Definition of the TKRS configuration grammar}
\label{t5}
{\fontsize{6pt}{7pt}\selectfont
\begin{tabular*}{\columnwidth}{p{15cm}}
\hline
\begin{verbatim}
 01  grammar org.xtext.TKRS.config.Config with org.eclipse.xtext.common.Terminals
 02
 03  import "platform:/resource/TKRS/metamodels/Configuration.ecore"
 04  import "platform:/resource/TKRS/metamodels/TKRS.ecore" as TKRSPackage
 05  import "platform:/resource/TKRS/metamodels/ImplementationRepository.ecore"
       as ImplementationRepositoryPackage
 06  import "http://www.eclipse.org/emf/2002/Ecore" as ecore
 07
 08  Package returns Package:
 09    {Package}
 10    'Package' name=EString
 11    ( hasImport+=Import (hasImport+=Import)* )? ( hasTKRS=TKRS )?
 12    ('ImplementationRepository' '{' hasImplementationRepository+=Platform
         (hasImplementationRepository+=Platform)* '}' )?
 13    ('Configuration' '{' hasStatement+=Statement (hasStatement+=Statement)* '}' )?;
 14
 15  Import returns Import: 'import' importedNamespace=ImportedFQN;
 16
 17  ImportedFQN: FQN ('.' '*')?;
 18
 19  FQN: ID ('.' ID)*;
 20
 21  TKRS returns TKRSPackage::TKRS: 'TKRS' name=EString '{' hasNode+=Node (hasNode+=Node)* '}';
 22
 23  Node returns TKRSPackage::Node:'Node' name=EString '{'
 24      'ip' ip=EString
 25      'port' port=EString
 26      'dbport' dbport=EString
 27      hasServiceModule=ServiceModule
 28      hasManagementModule=ManagementModule
 29      ( hasTradingModule+=TradingModule (hasTradingModule+=TradingModule)* )? 
 30            hasQueryModule+=QueryModule (hasQueryModule+=QueryModule)*
 31      ( hasProcessingModule+=ProcessingModule (hasProcessingModule+=ProcessingModule)* )? 
 32            'hasTKRS' hasTKRS=[TKRSPackage::TKRS|EString] '}';
 33
 34  TKRSModule returns TKRSPackage::Module:ManagementModule | QueryModule | TradingModule | ProcessingModule |
       ServiceModule;
 35
 36  ServiceModule returns TKRSPackage::ServiceModule:'ServiceModule' name=EString '{' 
 37          'hasNode' hasNode=[TKRSPackage::Node|EString] '}';
 38
 39  ManagementModule returns TKRSPackage::ManagementModule:'ManagementModule' name=EString '{' 
 40          'hasNode' hasNode=[TKRSPackage::Node|EString] '}';
 41
 42  QueryModule returns TKRSPackage::QueryModule: 'QueryModule' name=EString '{'
 43      'usesLookupInterface' usesLookupInterface=[TKRSPackage::TradingModule|EString]
 44      'hasNode' hasNode=[TKRSPackage::Node|EString] '}';
 45
 46  TradingModule returns TKRSPackage::TradingModule:'TradingModule' name=EString '{'
 47      'usesLookupInterface' lookupInterface=EBoolean
 48      'usesRegisterInterface' registerInterface=EBoolean
 49      'usesAdminInterface' adminInterface=EBoolean
 50      'usesLinkInterface' linkInterface=EBoolean
 51      'usesProxyInterface' proxyInterface=EBoolean
 52      ('isFederatedWith' isFederatedWith+=[TKRSPackage::TradingModule|EString]
 53      (isFederatedWith+=[TKRSPackage::TradingModule|EString])* )?
 54      'hasNode' hasNode=[TKRSPackage::Node|EString] '}';
 55
 56  ProcessingModule returns TKRSPackage::ProcessingModule:'ProcessingModule' name=EString '{'
 57      'ambient' ambient=EString
 58      'usesRegisterInterface' usesRegisterInterface=[TKRSPackage::TradingModule|EString]
 59      'hasNode' hasNode=[TKRSPackage::Node|EString] '}';
 60
 61  Platform returns ImplementationRepositoryPackage::Platform:'Platform' name=EString '{' 
 62       hasModule+=IRModule (hasModule+=IRModule)* '}';
 63
 64  IRModule returns ImplementationRepositoryPackage::Module: CompositeModule | SimpleModule;
 65
 66  CompositeModule returns ImplementationRepositoryPackage::CompositeModule:'CompositeModule' name=EString '{'
 67      'uri' uri=EString
 68      hasSubmodule+=IRModule (hasSubmodule+=IRModule)*
 69      ('hasPlatform' hasPlatform=[ImplementationRepositoryPackage::Platform|EString])?
 70      ('hasSuperModule' hasSuperModule=[ImplementationRepositoryPackage::Module|EString])? '}';
 71
 72   SimpleModule returns ImplementationRepositoryPackage::SimpleModule:'SimpleModule' name=EString '{'
 73      'uri' uri=EString
 74      ('hasPlatform' hasPlatform=[ImplementationRepositoryPackage::Platform|EString])?
 75      ('hasSuperModule' hasSuperModule=[ImplementationRepositoryPackage::Module|EString])? '}';
 76
 77  Statement returns Statement:'Statement '{'
 78      'hasTKRSModule' hasTKRSModule=[TKRSPackage::Module|EString]
 79      'hasImplementationRepositoryModule' hasImplementationRepositoryModule=
 80      [ImplementationRepositoryPackage::Module|EString] '}';
 81
 82  EString returns ecore::EString: STRING | ID;
 83  EBoolean returns ecore::EBoolean: 'true' | 'false';
 
\end{verbatim}
\\
\hline
\end{tabular*}
}
\end{table}

Table \ref{t5} shows the grammar defined for the configuration language. The grammar statement appears in line \#1, showing its location, and then, use of the \textit{org:eclipse:xtext:common:Terminals}, which defines rules for common terminals such as ID, STRING and INT. From line \#3 to \#6 ``.ecore'' files are declared with the metamodels used, and in the following lines, mapping between the text and the model language is defined. The input rule for the parser appears in line \#8. This rule shows that each Package entity begins with the reserved word Package followed by its name and the symbol ``\{''. In continuation, this entity defines a clause so the architecture models and implementation repository (PSM/M1 in Figure \ref{f1}) can be imported, although an M2M transformation must be applied to it before it can reference its elements from the configuration model, and continue with definition of all the elements in the three metamodels. It ends with the symbol ``\}''. As all the elements are defined the same way, one of them is given as a concrete example to make it easier to understand. The rule defined starting in line \#23 shows that each Node forming part of the TKRS begins with the reserved word Node followed by its name and the ``\{'' symbol. Then its attributes are included with another reserved word (such as \textit{ip}, \textit{port}, etc.), followed by the attribute \textit{value}. In continuation, this entity defines the modules that comprise it (with \textit{hasServiceModule}, \textit{hasManagementModule}, etc., relationships), referring to the system it operates in using the \textit{hasTKRS} relationship, and ends with the ``\}'' symbol. The system configuration models (PSM/M1) are created using an Eclipse editor. The SOLERES-KRS configuration model is shown in Section \ref{s63}.

It is already possible to deploy the system in a real environment based on the configuration model. To show an example of how this process can be automated, an M2T transformation has been implemented using the Eclipse Xpand tool, which enables the files necessary for a Java environment to be generated from the model created with the DSL editor. Table \ref{t6} shows the transformation template defined. This template describes the mapping between each entity of the model and its corresponding code. The sentence described starting in line \#5 refers to the Package entity. This definition generates a script file (\textit{make.sh}) in line \#8 containing operating system commands like \texttt{mkdir}, \texttt{cd} or \texttt{wget} to create a folder structure containing the necessary files. Line \#13 now generates a Java file called ``TKRS.java'' for each system node which is placed in a folder that receives its name from the value of the node attribute name, and which is, in turn, the subfolder of another with the value of the system attribute name. The content of these files is stated starting in line \#14. The attributes (lines \#20-\#38), check whether the node has each of the modules defined in the TKRS to include the corresponding attribute (for example, the QueryModule is found in lines \#23-\#29). The class constructor (lines \#40-\#60) is where the attributes defined above are initialized, etc. When the transformation has been executed, TKRS implementation is complete.

Once the TKRS framework has been described in terms of its (a) architecture, (b) modules of the trading service, (c) ontologies for using lookup, register and admin interfaces, (d) structure of data repository, and (e) configuration mechanisms, is important to formally define all these elements of the framework. Therefore, the following section provides the necessary definitions to describe the different parts of our approach, and is complemented by some algebraic expressions which help in such explanations.     

\begin{table}
\centering
\caption{M2T transformation template using Eclipse Xpand}
\label{t6}
{\fontsize{6pt}{7pt}\selectfont
\begin{tabular}{p{14cm}}
\hline
\vspace{-2mm}
\begin{alltt}
  1  {\guillemotleft}REM{\guillemotright}
  2    {\guillemotleft}IMPORT org::xtext::tkrs::config::config{\guillemotright}
  3  {\guillemotleft}ENDREM{\guillemotright}
  4
  5  {\guillemotleft}DEFINE Package FOR ConfigurationPackage::Package{\guillemotright}
  6    {\guillemotleft}IF !this.hasStatement.isEmpty{\guillemotright}
  7      {\guillemotleft}REM{\guillemotright} ------------------------------------------------------------------------ {\guillemotleft}ENDREM{\guillemotright}
  8      {\guillemotleft}FILE "make.sh"{\guillemotright}
  9        #!/bin/bash
 10        clear
 11        {\guillemotleft}FOREACH this.hasStatement.first().hasISModule.hasNode.hasTKRS.hasNode AS node{\guillemotright}
 12          {\guillemotleft}REM{\guillemotright} -------------------------------------------------------------------- {\guillemotleft}ENDREM{\guillemotright}
 13          {\guillemotleft}FILE node.hasTKRS.name + "/" + node.name + "/TKRS.java"{\guillemotright}
 14            package {\guillemotleft}node.hasTKRS.name{\guillemotright}.{\guillemotleft}node.name{\guillemotright};
 15
 16            import {\guillemotleft}node.hasTKRS.name{\guillemotright}.{\guillemotleft}node.name{\guillemotright}.modules.*;
 17
 18            public class TKRS \{
 20              private String ip = null;
 21              private int port = -1;
 22              private int dbport = -1;
 23              {\guillemotleft}FOREACH this.hasStatement AS statement{\guillemotright}
 24                {\guillemotleft}IF node==statement.hasTKRSModule.hasNode{\guillemotright}
 25                  {\guillemotleft}IF statement.hasTKRSModule.metaType.name=="TKRSPackage::QueryModule"{\guillemotright}
 26                    private QueryModule queryModule = null;
 27                  {\guillemotleft}ENDIF{\guillemotright}
 28                {\guillemotleft}ENDIF{\guillemotright}
 29              {\guillemotleft}ENDFOREACH{\guillemotright}
 30              {\guillemotleft}FOREACH this.hasStatement AS statement{\guillemotright}
 31                {\guillemotleft}IF node==statement.hasTKRSModule.hasNode{\guillemotright}
 32                  {\guillemotleft}IF statement.hasTKRSModule.metaType.name=="TKRSPackage::TradingModule"{\guillemotright}
 33                    private TradingModule tradingModule = null;
 34                  {\guillemotleft}ENDIF{\guillemotright}
 35                {\guillemotleft}ENDIF{\guillemotright}
 36              {\guillemotleft}ENDFOREACH{\guillemotright}
 37             ...
 38             // the same for ServiceModule, ManagementModule and ProcessingModule
 39
 40              public TKRS() \{
 41                this.ip = "{\guillemotleft}node.ip{\guillemotright}";
 42                this.port = {\guillemotleft}node.port{\guillemotright};
 43                this.dbport = {\guillemotleft}node.dbport{\guillemotright};
 44                {\guillemotleft}FOREACH this.hasStatement AS statement{\guillemotright}
 45                  {\guillemotleft}IF node==statement.hasTKRSModule.hasNode{\guillemotright}
 46                    {\guillemotleft}IF statement.hasTKRSModule.metaType.name=="TKRSPackage::QueryModule"{\guillemotright}
 47                      this.queryModule = new QueryModule();
 48                    {\guillemotleft}ENDIF{\guillemotright}
 49                  {\guillemotleft}ENDIF{\guillemotright}
 50                {\guillemotleft}ENDFOREACH{\guillemotright}
 51                {\guillemotleft}FOREACH this.hasStatement AS statement{\guillemotright}
 52                  {\guillemotleft}IF node==statement.hasTKRSModule.hasNode{\guillemotright}
 53                    {\guillemotleft}IF statement.hasTKRSModule.metaType.name=="TKRSPackage::TradingModule"{\guillemotright}
 54                      this.tradingModule = new TradingModule();
 55                    {\guillemotleft}ENDIF{\guillemotright}
 56                  {\guillemotleft}ENDIF{\guillemotright}
 57                {\guillemotleft}ENDFOREACH{\guillemotright}
 58             ...
 59             // the same for ServiceModule, ManagementModule and ProcessingModule
 60              \}
 61              // CODE
 62            \}
 63          {\guillemotleft}ENDFILE{\guillemotright}
 64          {\guillemotleft}REM{\guillemotright} -------------------------------------------------------------------- {\guillemotleft}ENDREM{\guillemotright}
 65          mkdir /{\guillemotleft}this.hasStatement.first().hasTKRSModule.hasNode.hasTKRS.name{\guillemotright}/{\guillemotleft}node.name{\guillemotright}/modules
 66          cd /{\guillemotleft}this.hasStatement.first().hasTKRSModule.hasNode.hasTKRS.name{\guillemotright}/{\guillemotleft}node.name{\guillemotright}/modules
 67          {\guillemotleft}FOREACH this.hasStatement AS statement{\guillemotright}
 68            {\guillemotleft}IF node==statement.hasTKRSModule.hasNode{\guillemotright}
 69              wget {\guillemotleft}statement.hasImplementationRepositoryModule.uri{\guillemotright}
 70            {\guillemotleft}ENDIF{\guillemotright}
 71          {\guillemotleft}ENDFOREACH{\guillemotright}
 72          cd /{\guillemotleft}this.hasStatement.first().hasTKRSModule.hasNode.hasTKRS.name{\guillemotright}/{\guillemotleft}node.name{\guillemotright}
 73          javac TKRS.java
 74        {\guillemotleft}ENDFOREACH{\guillemotright}
 75        cd /
 76      {\guillemotleft}ENDFILE{\guillemotright}
 77      {\guillemotleft}REM{\guillemotright} ------------------------------------------------------------------------ {\guillemotleft}ENDREM{\guillemotright}
 78    {\guillemotleft}ENDIF{\guillemotright}
 79  {\guillemotleft}ENDDEFINE{\guillemotright}
 
\end{alltt}
\\
\hline
\end{tabular}
}
\end{table}

\section{Formalization}
\label{sec:formalization}

{\color{black}This section formally describes the TKRS infrastructure using Descriptive Logic (DL) expressions. This notation allows us to define each of the parts and elements that must be taken into account to deploy a system based on large volumes of data (Big Data management).}

\begin{MyDef}[TKRS] A TKRS system $\mathcal{T}$ is determined by two finite sets $\mathcal{T} = \{\mathcal{H}_{\mathcal{T}}, \mathcal{N}\}$, where: (a) $\mathcal{H}_{\mathcal{T}}$ represents a set of properties $\mathcal{H}_{\mathcal{T}} = \{\mathcal{H}_{\mathcal{T}}^{1}\}$, where $\mathcal{H}_{\mathcal{T}}^{1}$ is the name of the system ({\it name} in the conceptual model), and (b) $\mathcal{N}$ represents a set of nodes ({\it Node}) $(\mathcal{N} \neq \varnothing)$.
\end{MyDef}

\begin{MyDef}[Node] A node $\mathcal{N}$ is in turn determined by another two finite sets $\mathcal{N} = \{\mathcal{H}_{\mathcal{N}}, \mathcal{M}\}$, where these sets are defined as follows:

\begin{enumerate}
\item[-] $\mathcal{H}_{\mathcal{N}}$ is a set of properties $\mathcal{H}_{\mathcal{N}} = \{\mathcal{H}_{\mathcal{N}}^{1}, \cdots, \mathcal{H}_{\mathcal{N}}^{l}\} ~|~l = \{1, \cdots, 4\}$, where: $\mathcal{H}_{\mathcal{N}}^{1}$ is the node \textit{name}, $\mathcal{H}_{\mathcal{N}}^{2}$ is an IP, $\mathcal{H}_{\mathcal{N}}^{3}$ is a communications port, and $\mathcal{H}_{\mathcal{N}}^{4}$ is a secondary port for configuration and access to the repository ({\it dbport}).

\item[-] $\mathcal{M}$ is a set of modules ({\it Module}) $\mathcal{M} = \{\mathcal{V}, \mathcal{G}, \mathcal{Q}, \mathcal{I}, \mathcal{E}\}$, where: $\mathcal{V}$ is a finite set of service modules $\mathcal{V} = \{\mathcal{V}_{1}, \cdots, \mathcal{V}_{o}\} ~|~o = \{1, \cdots, n\}$ $(\mathcal{V} \neq \varnothing)$, $\mathcal{G}$ is a finite set of management modules $\mathcal{G} = \{\mathcal{G}_{1}, \cdots, \mathcal{G}_{p}\} ~|~p = \{1, \cdots, n\}$ $(\mathcal{G} \neq \varnothing)$, $\mathcal{Q}$ is a finite set of query modules $\mathcal{Q} = \{\mathcal{Q}_{1}, \cdots, \mathcal{Q}_{q}\} ~|~q = \{1, \cdots, n\}$ $(\mathcal{Q} \neq \varnothing)$, $\mathcal{I}$ is a finite set of trading modules $\mathcal{I} = \{\mathcal{I}_{1}, \cdots, \mathcal{I}_{r}\} ~|~r = \{1, \cdots, n\}$, and $\mathcal{E}$ is a finite set of processing modules $\mathcal{E} = \{\mathcal{E}_{1}, \cdots, \mathcal{E}_{s}\} ~|~s = \{1, \cdots, n\}$.
\end{enumerate}
\end{MyDef}

\noindent
In the \textit{TKRS} system $\mathcal{T}$, it has to be given that  $\mathcal{I} \neq \varnothing$ and $\mathcal{E} \neq \varnothing$, at least in one of its nodes.

\begin{MyDef}[Service module] A service module $\mathcal{V}_{n}$ is identified by the property $\mathcal{H}_{\mathcal{V}}^{1}$, which represents the module {\it name}. The service module provides a series of services to the other modules, including registering modules and components, verifying state, etc., as defined below.
\end{MyDef}

\begin{MyDef}[Management module] A management module $\mathcal{G}_{n}$ is identified by the property $\mathcal{H}_{\mathcal{G}}^{1}$, which represents the module {\it name}. The management module acts as a hub for the user interface and the rest of the modules, enabling their configuration and being responsible for managing user demands.
\end{MyDef}

\begin{MyDef}[Query module] A query module $\mathcal{Q}_{n}$ is identified by the property $\mathcal{H}_{\mathcal{Q}}^{1}$, which represents the module {\it name}. The query module is concerned exclusively with the information queries made by users.
\end{MyDef}

\begin{MyDef}[Trading module] A trading module $\mathcal{I}_{n}$ is defined by the set of properties $\mathcal{I}_{n} = \{\mathcal{H}_{\mathcal{I}}^{1}, \cdots, \mathcal{H}_{\mathcal{I}}^{t}\} ~|~t = \{1, \cdots, 6\}$, and where: $\mathcal{H}_{\mathcal{I}}^{1}$ is the module {\it name}, $\mathcal{H}_{\mathcal{I}}^{2}$ is a property that shows whether the module implements the trading service's \textit{Lookup} interface, $\mathcal{H}_{\mathcal{I}}^{3}$ is a property which shows whether or not the module implements the trading service's mandatory \textit{Register} interface, $\mathcal{H}_{\mathcal{I}}^{4}$ is a property that shows whether the module implements the trading service's \textit{Admin} interface or not, $\mathcal{H}_{\mathcal{I}}^{5}$ is a property which shows whether or not the module implements the trading service's  \textit{Link} interface, and $\mathcal{H}_{\mathcal{I}}^{6}$ is a property that shows whether or not the module implements the trading service's \textit{Proxy} interface.
\end{MyDef}

The trading module enables search and location of system information, setting up a filter based on the query parameters. 

\begin{MyDef}[Processing module] A processing module $\mathcal{E}_{n}$ is defined by the pair of properties $\mathcal{E}_{n} = \{\mathcal{H}_{\mathcal{E}}^{1}, \mathcal{H}_{\mathcal{E}}^{2}\}$, where: $\mathcal{H}_{\mathcal{E}}^{1}$ is the module {\it name}, and $\mathcal{H}_{\mathcal{E}}^{2}$ is the name of the ambient the module belongs to, since processing modules can be grouped by different criteria, such as their location, the information they manage, etc. 
\end{MyDef}

The processing module is responsible for managing knowledge sources. 

\begin{MyDef}[Query Relationship] Given a query module $\mathcal{Q}_{n}$ and a trading module $\mathcal{I}_{n}$, the query relationship $\mathcal{R}_{\mathcal{K}}$ is defined as $\mathcal{R}_{\mathcal{K}} \subseteq \mathcal{Q}_{n} \times \mathcal{I}_{n}$. Its purpose is to define the relationship between the two modules for information queries.
\end{MyDef}

\begin{MyDef}[Register relationship] Given a processing module $\mathcal{E}_{n}$ and a trading module $\mathcal{I}_{n}$, the register relationship $\mathcal{R}_{\mathcal{G}}$ is defined as $\mathcal{R}_{\mathcal{G}} \subseteq \mathcal{E}_{n} \times \mathcal{I}_{n}$. Its purpose consists of defining the relationship between the two modules for information management. 
\end{MyDef}

\begin{MyDef}[Federation relationship] Given two trading modules $\mathcal{I}_{n}$ and $\mathcal{I}_{m}$ $(\mathcal{I}_{n} \neq \mathcal{I}_{m})$, the federation relationship $\mathcal{R}_{\mathcal{F}}$ is defined as $\mathcal{R}_{\mathcal{F}} \subseteq \mathcal{I}_{n} \times \mathcal{I}_{m}$.  Its purpose consists of defining the relationship between the two modules for the federation. The interfaces \textit{Link} of the modules must be active ({\it linkInterface = true}).
\end{MyDef}

\begin{MyDef}[Service/process ontologies] Service/process ontologies $\mathcal{S}$ are defined as a finite set $\mathcal{S} = \{\mathcal{C}, \mathcal{A}, \mathcal{P}\}$, where: $\mathcal{C}$ is a finite set of concepts, $\mathcal{C} = \{c_{1}, \cdots, c_{i}\} ~|~i = \{1, \cdots, n\} ~,~ n \in \cal{N}$; $\mathcal{A}$ is a finite set of actions, $\mathcal{A} = \{a_{1}, \cdots, a_{j}\} ~|~j = \{1, \cdots, n\} ~,~ n \in \cal{N}$; and $\mathcal{P}$ is a finite set of predicates, $\mathcal{P} = \{p_{1}, \cdots, p_{k}\} ~|~k = \{1, \cdots, n\} ~,~ n \in \cal{N}$.
\end{MyDef}

The service/process ontologies give the possible actions $\mathcal{A}$ that can be taken by means of each trading service interface, the information $\mathcal{C}$ necessary to perform them and the result P when completed (either a controlled exception during the action or else that it has been performed satisfactorily). 

Given a set of concepts $\mathcal{C}_{n} \mid \mathcal{C}_{n} \subset \mathcal{C}$, a set of actions  $\mathcal{A}_{n} \mid \mathcal{A}_{n} \subset \mathcal{A}$ and a set of predicates $\mathcal{P}_{n} \mid \mathcal{P}_{n} \subset \mathcal{P}$, then:

\begin{enumerate} 
	\item[(i)] $\mathcal{C}_{n} \times \mathcal{A}_{n} \times \mathcal{P}_{n} \Rightarrow \mathcal{S}_{n}$
\end{enumerate}

\noindent where $\mathcal{S}_{n}$ is a concrete ontology of the five associated with the different trading service interfaces: \textit{Lookup}, \textit{Register}, \textit{Admin}, \textit{Link} and \textit{Proxy}.

A trader should have at least the functionality of the \textit{Lookup} and \textit{Register} interfaces, along with their corresponding ontologies: the first to enable documents stored in the repository to be queried, and the second to manage these documents. The presence of the rest of the interfaces is not obligatory: the \textit{Admin} ontology would facilitate configuration of trader parameters and policies (for example, number of federated traders allowed, search policies, etc.); the \textit{Link} ontology refers to how the traders interconnect (federation process); and finally, the \textit{Proxy} ontology models old properties in trading federation systems. 

The formalization of each of the ontologies related to the corresponding stand-alone trading service model interface described in Section \ref{sec:framework} is shown below. 

\begin{MyDef}[\textit{Lookup} Ontology] Given the set of concepts $\mathcal{C}_{\mathcal{L}}$, the set of actions $\mathcal{A}_{\mathcal{L}}$ and the set of predicates $\mathcal{P}_{\mathcal{L}}$, all of them referring to the \textit{Lookup} ontology $\mathcal{S}_{\mathcal{L}}$, this ontology is defined as $\mathcal{C}_{\mathcal{L}} \times \mathcal{A}_{\mathcal{L}} \times \mathcal{P}_{\mathcal{L}} \Rightarrow \mathcal{S}_{\mathcal{L}}$, where:

\begin{enumerate} 
	\item[(i)] $\mathcal{C}_{\mathcal{L}} = \{\mathcal{C}_{\mathcal{L}}^{1}, \cdots, \mathcal{C}_{\mathcal{L}}^{i}\} ~|~i = \{1, \cdots, 4\}$, where: $\mathcal{C}_{\mathcal{L}}^{1}$ is a query ({\it QueryForm} in the conceptual model), $\mathcal{C}_{\mathcal{L}}^{2}$ is a sequence of policies to be followed with the query ({\it PolicySeq}), $\mathcal{C}_{\mathcal{L}}^{3}$ is a sequence of results ({\it OfferSeq}), $\mathcal{C}_{\mathcal{L}}^{4}$ is a sequence with the percentage of coincidences of each result ({\it OfferSeqMatch}).

	\item[(ii)] $\mathcal{A}_{\mathcal{L}} = \{\mathcal{A}_{\mathcal{L}}^{1}\}$, where: $\mathcal{A}_{\mathcal{L}}^{1}$ is the action of consulting metadata ({\it Query}).

	\item[(iii)] $\mathcal{P}_{\mathcal{L}} = \{\mathcal{P}_{\mathcal{L}}^{1}, \cdots, \mathcal{P}_{\mathcal{L}}^{k}\} ~|~k = \{1, \cdots, 7\}$, where: $\mathcal{P}_{\mathcal{L}}^{1}, \cdots, \mathcal{P}_{\mathcal{L}}^{5}$ is a controlled exception during the actions of set $\mathcal{A}_{\mathcal{L}}$ ({\it UnknownQueryForm}, {\it PolicyTypeMismatch}, {\it InvalidPolicyValue}, {\it DuplicatePolicyName} and {\it QueryError}, respectively); and $\mathcal{P}_{\mathcal{L}}^{6}$ and $\mathcal{P}_{\mathcal{L}}^{7}$ are actions $\mathcal{A}_{\mathcal{L}}$ performed satisfactorily ({\it EmptyOfferSeq} and {\it NotEmptyOfferSeq}, respectively).
\end{enumerate}
\end{MyDef}

\begin{MyDef}[Request relationship -- \textit{Lookup} ontology] Given the concepts $\mathcal{C}_{\mathcal{L}}^{1}$ and $\mathcal{C}_{\mathcal{L}}^{2}$, and action $\mathcal{A}_{\mathcal{L}}^{1}$ in the \textit{Lookup} $\mathcal{S}_{\mathcal{L}}$ ontology, the relationship of request $\mathcal{R}_{\mathcal{L}}$ is defined as $\mathcal{R}_{\mathcal{L}} \subseteq (\mathcal{C}_{\mathcal{L}}^{1} \times \mathcal{A}_{\mathcal{L}}^{1}) \cup (\mathcal{C}_{\mathcal{L}}^{1} \times \mathcal{A}_{\mathcal{L}}^{1} \times \mathcal{C}_{\mathcal{L}}^{2})$.
\end{MyDef}

\begin{MyDef}[Response relationship -- \textit{Lookup} ontology] Given an action $\mathcal{A}_{\mathcal{L}}^{1}$, a predicate $\mathcal{P}_{\mathcal{L}}^{k} ~|~k = \{1, \cdots, 7\}$ and concepts $\mathcal{C}_{\mathcal{L}}^{3}$ and $\mathcal{C}_{\mathcal{L}}^{4}$ in \textit{Lookup} $\mathcal{S}_{\mathcal{L}}$ ontology, the relationship of response $\overline{\mathcal{R}_{\mathcal{L}}}$ is defined as $\overline{\mathcal{R}_{\mathcal{L}}} \subseteq (\mathcal{A}_{\mathcal{L}}^{1} \times \mathcal{P}_{\mathcal{L}}^{k}) \cup (\mathcal{A}_{\mathcal{L}}^{1} \times \mathcal{P}_{\mathcal{L}}^{k} \times \mathcal{C}_{\mathcal{L}}^{3}) \cup (\mathcal{A}_{\mathcal{L}}^{1} \times \mathcal{P}_{\mathcal{L}}^{k} \times \mathcal{C}_{\mathcal{L}}^{3} \times \mathcal{C}_{\mathcal{L}}^{4})$.
\end{MyDef}

\begin{MyDef}[Uses of the \textit{Lookup} ontology] Given action $\mathcal{A}_{\mathcal{L}}^{1}$, a predicate $\mathcal{P}_{\mathcal{L}}^{k} ~|~k = \{1, \cdots, 7\}$ and concepts $\mathcal{C}_{\mathcal{L}}^{3}$ and $\mathcal{C}_{\mathcal{L}}^{4}$ in the \textit{Lookup} $\mathcal{S}_{\mathcal{L}}$ ontology, the following conditions $f(\mathcal{S}_{\mathcal{L}})$ hold true for its uses:

\begin{equation}
f(S_L) = \left\{
\begin{array}{l}
(\mathcal{A}_{\mathcal{L}}^{1} \times \mathcal{P}_{\mathcal{L}}^{k}) \Rightarrow \mathcal{P}_{\mathcal{L}}^{k} = \mathcal{P}_{\mathcal{L}}^{1} \vee \mathcal{P}_{\mathcal{L}}^{2} \vee \mathcal{P}_{\mathcal{L}}^{3} \vee \mathcal{P}_{\mathcal{L}}^{4} \vee \mathcal{P}_{\mathcal{L}}^{5} \vee \mathcal{P}_{\mathcal{L}}^{6}\\
	
(\mathcal{A}_{\mathcal{L}}^{1} \times \mathcal{P}_{\mathcal{L}}^{k} \times \mathcal{C}_{\mathcal{L}}^{3}) \Rightarrow \mathcal{P}_{\mathcal{L}}^{k} = \mathcal{P}_{\mathcal{L}}^{7}\\
	
(\mathcal{A}_{\mathcal{L}}^{1} \times \mathcal{P}_{\mathcal{L}}^{k} \times \mathcal{C}_{\mathcal{L}}^{3} \times \mathcal{C}_{\mathcal{L}}^{4}) \Rightarrow \mathcal{P}_{\mathcal{L}}^{k} = \mathcal{P}_{\mathcal{L}}^{7}\\
\end{array}
\right.
\end{equation}
\end{MyDef}

\begin{MyDef}[\textit{Register} Ontology] Given the set of concepts $\mathcal{C}_{\mathcal{R}}$, set of actions $\mathcal{A}_{\mathcal{R}}$ and set of predicates $\mathcal{P}_{\mathcal{R}}$, all of them referring to \textit{Register} ontology $\mathcal{S}_{\mathcal{R}}$, this ontology is defined as $\mathcal{C}_{\mathcal{R}} \times \mathcal{A}_{\mathcal{R}} \times \mathcal{P}_{\mathcal{R}} \Rightarrow \mathcal{S}_{\mathcal{R}}$, where:

\begin{enumerate}
	\item[(i)] $\mathcal{C}_{\mathcal{R}} = \{\mathcal{C}_{\mathcal{R}}^{1}, \mathcal{C}_{\mathcal{R}}^{2}\}$, where: $\mathcal{C}_{\mathcal{R}}^{1}$ is the \textit{URI} of a metadata document ({\it Offer} in the conceptual model); $\mathcal{C}_{\mathcal{R}}^{2}$ is the unique identifier of a metadata document ({\it OfferId}).

	\item[(ii)] $\mathcal{A}_{\mathcal{R}} = \{\mathcal{A}_{\mathcal{R}}^{1}, \cdots, \mathcal{A}_{\mathcal{R}}^{j}\} ~|~j = \{1, \cdots, 4\}$, where: $\mathcal{A}_{\mathcal{R}}^{1}$ is the action of adding a metadata document ({\it Export}); $\mathcal{A}_{\mathcal{R}}^{2}$ represents the action of modifying a metadata document ({\it Modify}); $\mathcal{A}_{\mathcal{R}}^{3}$ is the action of eliminating a metadata document ({\it Withdraw}); and $\mathcal{A}_{\mathcal{R}}^{4}$ is the action of querying a metadata document ({\it Describe}).

	\item[(iii)] $\mathcal{P}_{\mathcal{R}} = \{\mathcal{P}_{\mathcal{R}}^{1}, \cdots, \mathcal{P}_{\mathcal{R}}^{k}\} ~|~k = \{1, \cdots, 11\}$, where: $\mathcal{P}_{\mathcal{R}}^{1}, \cdots, \mathcal{P}_{\mathcal{R}}^{6}$ are controlled exceptions during set of actions $\mathcal{A}_{\mathcal{R}}$ ({\it IllegalOffer}, {\it UnknownOffer}, {\it DuplicateOffer}, {\it IllegalOfferId}, {\it UnknownOfferId} and {\it QueryError}, respectively); $\mathcal{P}_{\mathcal{R}}^{7}, \cdots, \mathcal{P}_{\mathcal{R}}^{11}$ show that actions $\mathcal{A}_{\mathcal{R}}$ have been performed satisfactorily ({\it Exported-}{\it Offer}, {\it WithdrawnOffer}, {\it DescribedOffer}, {\it NotDescribedOffer} and {\it Modified-}{\it Offer}, respectively).
\end{enumerate}
\end{MyDef}

\begin{MyDef}[Request relationship -- \textit{Register} ontology] Given the concepts $\mathcal{C}_{\mathcal{R}}^{1}$ and $\mathcal{C}_{\mathcal{R}}^{2}$, and actions $\mathcal{A}_{\mathcal{R}}^{1}, \cdots, \mathcal{A}_{\mathcal{R}}^{4}$ in \textit{Register} ontology $\mathcal{S}_{\mathcal{R}}$, the relationship of request $\mathcal{R}_{\mathcal{R}}$ is defined as $\mathcal{R}_{\mathcal{R}} \subseteq (\mathcal{C}_{\mathcal{R}}^{1} \times \mathcal{A}_{\mathcal{R}}^{1}) \cup (\mathcal{C}_{\mathcal{R}}^{2} \times \mathcal{C}_{\mathcal{R}}^{1} \times \mathcal{A}_{\mathcal{R}}^{2}) \cup (\mathcal{C}_{\mathcal{R}}^{2} \times \mathcal{A}_{\mathcal{R}}^{3}) \cup (\mathcal{C}_{\mathcal{R}}^{2} \times \mathcal{A}_{\mathcal{R}}^{4})$.
\end{MyDef}

\label{def:register}
\begin{MyDef}[Response relationship -- \textit{Register} ontology] Given the actions $\mathcal{A}_{\mathcal{R}}^{1}, \cdots, \mathcal{A}_{\mathcal{R}}^{4}$, a predicate $\mathcal{P}_{\mathcal{R}}^{k} ~|~k = \{1, \cdots, 11\}$ and concepts $\mathcal{C}_{\mathcal{R}}^{1}$ and $\mathcal{C}_{\mathcal{R}}^{2}$ in \textit{Register} ontology $\mathcal{S}_{\mathcal{R}}$, the relationship of response $\overline{\mathcal{R}_{\mathcal{R}}}$ is defined as $\overline{\mathcal{R}_{\mathcal{R}}} \subseteq (\mathcal{A}_{\mathcal{R}}^{1} \times \mathcal{P}_{\mathcal{R}}^{k}) \cup (\mathcal{A}_{\mathcal{R}}^{2} \times \mathcal{P}_{\mathcal{R}}^{k}) \cup (\mathcal{A}_{\mathcal{R}}^{3} \times \mathcal{P}_{\mathcal{R}}^{k}) \cup (\mathcal{A}_{\mathcal{R}}^{4} \times \mathcal{P}_{\mathcal{R}}^{k}) \cup (\mathcal{A}_{\mathcal{R}}^{1} \times \mathcal{P}_{\mathcal{R}}^{k} \times \mathcal{C}_{\mathcal{R}}^{2}) \cup (\mathcal{A}_{\mathcal{R}}^{2} \times \mathcal{P}_{\mathcal{R}}^{k} \times \mathcal{C}_{\mathcal{R}}^{2}) \cup (\mathcal{A}_{\mathcal{R}}^{3} \times \mathcal{P}_{\mathcal{R}}^{k} \times \mathcal{C}_{\mathcal{R}}^{1}) \cup (\mathcal{A}_{\mathcal{R}}^{4} \times \mathcal{P}_{\mathcal{R}}^{k} \times \mathcal{C}_{\mathcal{R}}^{1})$.
\end{MyDef}

\begin{MyDef}[Uses of the \textit{Register} ontology] Given actions $\mathcal{A}_{\mathcal{R}}^{1}, \cdots, \mathcal{A}_{\mathcal{R}}^{4}$, a predicate $\mathcal{P}_{\mathcal{R}}^{k} ~|~ k = \{1, \cdots, 11\}$ (defined in Definition 16.iii) and concepts $\mathcal{C}_{\mathcal{R}}^{1}$ and $\mathcal{C}_{\mathcal{R}}^{2}$ in \textit{Register} ontology $\mathcal{S}_{\mathcal{R}}$,the following conditions $f(\mathcal{S}_{\mathcal{R}})$ hold true for its uses:

\begin{equation}
f(S_R) = \left\{
\begin{array}{l}
(\mathcal{A}_{\mathcal{R}}^{1} \times \mathcal{P}_{\mathcal{R}}^{k}) \Rightarrow \mathcal{P}_{\mathcal{R}}^{k} = \mathcal{P}_{\mathcal{R}}^{4} \vee \mathcal{P}_{\mathcal{R}}^{2} \vee \mathcal{P}_{\mathcal{R}}^{3}\\
		
(\mathcal{A}_{\mathcal{R}}^{2} \times \mathcal{P}_{\mathcal{R}}^{k}) \Rightarrow \mathcal{P}_{\mathcal{R}}^{k} = \mathcal{P}_{\mathcal{R}}^{4} \vee \mathcal{P}_{\mathcal{R}}^{5} \vee \mathcal{P}_{\mathcal{R}}^{1} \vee \mathcal{P}_{\mathcal{R}}^{2} \vee \mathcal{P}_{\mathcal{R}}^{3}\\
			
(\mathcal{A}_{\mathcal{R}}^{3} \times \mathcal{P}_{\mathcal{R}}^{k}) \Rightarrow \mathcal{P}_{\mathcal{R}}^{k} = \mathcal{P}_{\mathcal{R}}^{4} \vee \mathcal{P}_{\mathcal{R}}^{5}\\
		
(\mathcal{A}_{\mathcal{R}}^{4} \times \mathcal{P}_{\mathcal{R}}^{k}) \Rightarrow \mathcal{P}_{\mathcal{R}}^{k} = \mathcal{P}_{\mathcal{R}}^{6} \vee \mathcal{P}_{\mathcal{R}}^{4} \vee \mathcal{P}_{\mathcal{R}}^{10}\\
		
(\mathcal{A}_{\mathcal{R}}^{1} \times \mathcal{P}_{\mathcal{R}}^{k} \times \mathcal{C}_{\mathcal{R}}^{2}) \Rightarrow \mathcal{P}_{\mathcal{R}}^{k} = \mathcal{P}_{\mathcal{R}}^{7}\\
		
(\mathcal{A}_{\mathcal{R}}^{2} \times \mathcal{P}_{\mathcal{R}}^{k} \times \mathcal{C}_{\mathcal{R}}^{2}) \Rightarrow \mathcal{P}_{\mathcal{R}}^{k} = \mathcal{P}_{\mathcal{R}}^{11}\\
		
(\mathcal{A}_{\mathcal{R}}^{3} \times \mathcal{P}_{\mathcal{R}}^{k} \times \mathcal{C}_{\mathcal{R}}^{1}) \Rightarrow \mathcal{P}_{\mathcal{R}}^{k} = \mathcal{P}_{\mathcal{R}}^{8}\\
		
(\mathcal{A}_{\mathcal{R}}^{4} \times \mathcal{P}_{\mathcal{R}}^{k} \times \mathcal{C}_{\mathcal{R}}^{1}) \Rightarrow \mathcal{P}_{\mathcal{R}}^{k} = \mathcal{P}_{\mathcal{R}}^{9}\\
\end{array}
\right.
\end{equation}
\end{MyDef}

\begin{MyDef}[\textit{Admin} Ontology] Given set of concepts $\mathcal{C}_{\mathcal{A}}$, set of actions $\mathcal{A}_{\mathcal{A}}$ and set of predicates $\mathcal{P}_{\mathcal{A}}$, all referring to \textit{Admin} $\mathcal{S}_{\mathcal{A}}$, this ontology is defined as $\mathcal{C}_{\mathcal{A}} \times \mathcal{A}_{\mathcal{A}} \times \mathcal{P}_{\mathcal{A}} \Rightarrow \mathcal{S}_{\mathcal{A}}$, where:

\begin{enumerate} 
	\item[(i)] $\mathcal{C}_{\mathcal{A}} = \{\mathcal{C}_{\mathcal{A}}^{1}, \mathcal{C}_{\mathcal{A}}^{2}, \mathcal{C}_{\mathcal{A}}^{3}\}$, where: $\mathcal{C}_{\mathcal{A}}^{1}$ is the default number of documents to be located in a search ({\it Def\_search\_card} in the conceptual model); $\mathcal{C}_{\mathcal{A}}^{2}$ is the maximum number of documents to be located in a search ({\it Max\_search\_card}); $\mathcal{C}_{\mathcal{A}}^{3}$ is the \textit{URI} of the repository with the metadata documents used by the trading service ({\it Offer\_repos}).

	\item[(ii)] $\mathcal{A}_{\mathcal{A}} = \{\mathcal{A}_{\mathcal{A}}^{1}, \cdots, \mathcal{A}_{\mathcal{A}}^{j}\} ~|~j = \{1, \cdots, 6\}$, where: $\mathcal{A}_{\mathcal{A}}^{1}$  is the action of setting the default number of documents to be located in a search ({\it SetDef\_search\_card}); $\mathcal{A}_{\mathcal{A}}^{2}$ is the action of setting the maximum number of documents to be located in a search ({\it SetMax\_search\_card}); $\mathcal{A}_{\mathcal{A}}^{3}$ is the action of setting the \textit{URI} of the repository with the metadata documents used by the trading service ({\it SetOffer\_repos}); $\mathcal{A}_{\mathcal{A}}^{4}$ is the action of returning the default number of documents to be located in a search ({\it GetDef\_search\_card}); $\mathcal{A}_{\mathcal{A}}^{5}$ is the action of returning the maximum number of documents to be located in a search ({\it GetMax\_search\_card}); and $\mathcal{A}_{\mathcal{A}}^{6}$ is the action of returning the \textit{URI} of the repository with the metadata documents used by the trading service ({\it GetOffer\_repos}).

	\item[(iii)] $\mathcal{P}_{\mathcal{A}} = \{\mathcal{P}_{\mathcal{A}}^{1}, \cdots, \mathcal{P}_{\mathcal{A}}^{k}\} ~|~k = \{1, \cdots, 7\}$, where: $\mathcal{P}_{\mathcal{A}}^{1}$ is a controlled exception during the actions in set $\mathcal{A}_{\mathcal{A}}$ ({\it InvalidValue}); $\mathcal{P}_{\mathcal{A}}^{2}, \cdots, \mathcal{P}_{\mathcal{A}}^{7}$ are actions $\mathcal{A}_{\mathcal{A}}$ performed satisfactorily ({\it ModifiedDef\_search\_card}, {\it ModifiedMax\_search\_card}, {\it ModifiedOffer\_repos}, {\it ReturnedDef\_search\_card}, {\it ReturnedMax\_search\_card} and {\it ReturnedOffer\_repos}, respectively).
\end{enumerate}
\end{MyDef}

\begin{MyDef}[Request relationship -- \textit{Admin} ontology] Given concepts $\mathcal{C}_{\mathcal{A}}^{1}, \cdots, \mathcal{C}_{\mathcal{A}}^{3}$, and actions $\mathcal{A}_{\mathcal{A}}^{1}, \cdots, \mathcal{A}_{\mathcal{A}}^{6}$ in the \textit{Admin} ontology $\mathcal{S}_{\mathcal{A}}$, the relationship of request $\mathcal{R}_{\mathcal{A}}$ is defined as $\mathcal{R}_{\mathcal{A}} \subseteq (\mathcal{C}_{\mathcal{A}}^{1} \times \mathcal{A}_{\mathcal{A}}^{1}) \cup (\mathcal{C}_{\mathcal{A}}^{2} \times \mathcal{A}_{\mathcal{A}}^{2}) \cup (\mathcal{C}_{\mathcal{A}}^{3} \times \mathcal{A}_{\mathcal{A}}^{3}) \cup (\mathcal{A}_{\mathcal{A}}^{4} \times \mathcal{C}_{\mathcal{A}}^{1}) \cup (\mathcal{A}_{\mathcal{A}}^{5} \times \mathcal{C}_{\mathcal{A}}^{2}) \cup (\mathcal{A}_{\mathcal{A}}^{6} \times \mathcal{C}_{\mathcal{A}}^{3})$.
\end{MyDef}

\begin{MyDef}[Response relationship -- \textit{Admin} ontology] Given actions $\mathcal{A}_{\mathcal{A}}^{1}, \cdots, \mathcal{A}_{\mathcal{A}}^{6}$, a predicate $\mathcal{P}_{\mathcal{A}}^{k} ~|~k = \{1, \cdots, 7\}$ and concepts $\mathcal{C}_{\mathcal{A}}^{1}, \cdots, \mathcal{C}_{\mathcal{A}}^{3}$ in \textit{Admin} ontology $\mathcal{S}_{\mathcal{A}}$, the relationship of response $\overline{\mathcal{R}_{\mathcal{A}}}$ is defined as $\overline{\mathcal{R}_{\mathcal{A}}} \subseteq (\mathcal{A}_{\mathcal{A}}^{1} \times \mathcal{P}_{\mathcal{A}}^{k} \times \mathcal{C}_{\mathcal{A}}^{1}) \cup (\mathcal{A}_{\mathcal{A}}^{2} \times \mathcal{P}_{\mathcal{A}}^{k} \times \mathcal{C}_{\mathcal{A}}^{2}) \cup (\mathcal{A}_{\mathcal{A}}^{3} \times \mathcal{P}_{\mathcal{A}}^{k} \times \mathcal{C}_{\mathcal{A}}^{3}) \cup (\mathcal{A}_{\mathcal{A}}^{4} \times \mathcal{P}_{\mathcal{A}}^{k} \times \mathcal{C}_{\mathcal{A}}^{1}) \cup (\mathcal{A}_{\mathcal{A}}^{5} \times \mathcal{P}_{\mathcal{A}}^{k} \times \mathcal{C}_{\mathcal{A}}^{2}) \cup (\mathcal{A}_{\mathcal{A}}^{6} \times \mathcal{P}_{\mathcal{A}}^{k} \times \mathcal{C}_{\mathcal{A}}^{3}) \cup (\mathcal{A}_{\mathcal{A}}^{1} \times \mathcal{P}_{\mathcal{A}}^{k}) \cup (\mathcal{A}_{\mathcal{A}}^{2} \times \mathcal{P}_{\mathcal{A}}^{k})$.
\end{MyDef}

\begin{MyDef}[Uses of the \textit{Admin} ontology] Given actions $\mathcal{A}_{\mathcal{A}}^{1}, \cdots, \mathcal{A}_{\mathcal{A}}^{6}$, a predicate $\mathcal{P}_{\mathcal{A}}^{k} ~|~k = \{1, \cdots, 7\}$ and concepts $\mathcal{C}_{\mathcal{A}}^{1}, \cdots, \mathcal{C}_{\mathcal{A}}^{3}$ in the \textit{Admin} ontology $\mathcal{S}_{\mathcal{A}}$, the following conditions $f(\mathcal{S}_{\mathcal{A}})$ hold true for its uses:

\begin{equation}
     f(S_A) = \left\{
	       \begin{array}{l}
(\mathcal{A}_{\mathcal{A}}^{1} \times \mathcal{P}_{\mathcal{A}}^{k} \times \mathcal{C}_{\mathcal{A}}^{1}) \Rightarrow \mathcal{P}_{\mathcal{A}}^{k} = \mathcal{P}_{\mathcal{A}}^{2}\\
(\mathcal{A}_{\mathcal{A}}^{2} \times \mathcal{P}_{\mathcal{A}}^{k} \times \mathcal{C}_{\mathcal{A}}^{2}) \Rightarrow \mathcal{P}_{\mathcal{A}}^{k} = \mathcal{P}_{\mathcal{A}}^{3}\\
(\mathcal{A}_{\mathcal{A}}^{3} \times \mathcal{P}_{\mathcal{A}}^{k} \times \mathcal{C}_{\mathcal{A}}^{3}) \Rightarrow \mathcal{P}_{\mathcal{A}}^{k} = \mathcal{P}_{\mathcal{A}}^{4}\\
(\mathcal{A}_{\mathcal{A}}^{4} \times \mathcal{P}_{\mathcal{A}}^{k} \times \mathcal{C}_{\mathcal{A}}^{1}) \Rightarrow \mathcal{P}_{\mathcal{A}}^{k} = \mathcal{P}_{\mathcal{A}}^{5}\\
(\mathcal{A}_{\mathcal{A}}^{5} \times \mathcal{P}_{\mathcal{A}}^{k} \times \mathcal{C}_{\mathcal{A}}^{2}) \Rightarrow \mathcal{P}_{\mathcal{A}}^{k} = \mathcal{P}_{\mathcal{A}}^{6}\\
(\mathcal{A}_{\mathcal{A}}^{6} \times \mathcal{P}_{\mathcal{A}}^{k} \times \mathcal{C}_{\mathcal{A}}^{3}) \Rightarrow \mathcal{P}_{\mathcal{A}}^{k} = \mathcal{P}_{\mathcal{A}}^{7}\\
(\mathcal{A}_{\mathcal{A}}^{1} \times \mathcal{P}_{\mathcal{A}}^{k}) \Rightarrow \mathcal{P}_{\mathcal{A}}^{k} = \mathcal{P}_{\mathcal{A}}^{1}\\
(\mathcal{A}_{\mathcal{A}}^{2} \times \mathcal{P}_{\mathcal{A}}^{k}) \Rightarrow \mathcal{P}_{\mathcal{A}}^{k} = \mathcal{P}_{\mathcal{A}}^{1}\\
\end{array}
\right.
\end{equation}

\end{MyDef}

\begin{MyDef}[Platform] A platform is determined by the pair $\mathcal{L} = \{\mathcal{H}_{\mathcal{L}}, \mathcal{W}\}$, where:
 
\begin{enumerate} 
	\item[(i)] $\mathcal{H}_{\mathcal{L}}$ is the property $\mathcal{H}_{\mathcal{L}}^{1}$ which gives the name of the platform ({\it name}).

	\item[(ii)] $\mathcal{W}$ is a set of module implementations ({\it Module}) $\mathcal{W} = \{\mathcal{X}, \mathcal{Y}\}$ $(\mathcal{W} \neq \varnothing)$, where: $\mathcal{X}$ is a finite set of composite module implementations in which $\mathcal{X} = \{\mathcal{X}_{1}, \cdots, \mathcal{X}_{u}\} ~|~u = \{1, \cdots, n\}$; and $\mathcal{Y}$ is a finite set of simple module implementations $\mathcal{Y} = \{\mathcal{Y}_{1}, \cdots, \mathcal{Y}_{v}\} ~|~v = \{1, \cdots, n\}$.
\end{enumerate}
\end{MyDef}

\begin{MyDef}[Composite module implementation] An implementation of a composite module $\mathcal{X}_{u}$ is defined in turn by three finite sets $\mathcal{X}_{u} = \{\mathcal{H}_{\mathcal{X}}, \mathcal{X}, \mathcal{Y}\}$, where:
 
\begin{enumerate} 
	\item[(i)] $\mathcal{H}_{\mathcal{X}}$ is the pair of properties $\mathcal{H}_{\mathcal{X}} = \{\mathcal{H}_{\mathcal{X}}^{1}, \mathcal{H}_{\mathcal{X}}^{2}\}$, where: $\mathcal{H}_{\mathcal{X}}^{1}$ is the name of the module implementation ({\it name}); and  $\mathcal{H}_{\mathcal{X}}^{2}$ is the location of the implementation files ({\it uri}).

	\item[(ii)] $\mathcal{X}$ is a finite set of implementations of composite modules $\mathcal{X} = \{\mathcal{X}_{1}, \cdots, \mathcal{X}_{u}\} ~|~u = \{1, \cdots, n\}$.

	\item[(iii)] $\mathcal{Y}$ is a finite set of simple implementation modules $\mathcal{Y} = \{\mathcal{Y}_{1}, \cdots, \mathcal{Y}_{v}\} ~|~v = \{1, \cdots, n\}$.
\end{enumerate}
\end{MyDef}

\begin{MyDef}[Simple module implementation] An implementation of a simple module ({\it SimpleModule} in the conceptual model) $\mathcal{Y}_{v}$ is defined by a finite set $\mathcal{Y}_{v} = \{\mathcal{H}_{\mathcal{Y}}\}$, where: $\mathcal{H}_{\mathcal{Y}}$ is a set of properties $\mathcal{H}_{\mathcal{Y}} = \{\mathcal{H}_{\mathcal{Y}}^{1}, \mathcal{H}_{\mathcal{Y}}^{2}\}$, and where: $\mathcal{H}_{\mathcal{Y}}^{1}$ is the name of the implementation module ({\it name}); and $\mathcal{H}_{\mathcal{Y}}^{2}$ is the location of the implementation files ({\it uri}).
\end{MyDef}

\begin{MyDef}[Platform relationship] Given an implementation of a composite module $\mathcal{X}_{u}$ or of a simple module $\mathcal{Y}_{v}$, and a platform $\mathcal{L}$, the relationship of platform $\mathcal{R}_{\mathcal{P}}$ is defined as $\mathcal{R}_{\mathcal{P}} \subseteq (\mathcal{L} \times \mathcal{X}_{u}) \vee (\mathcal{L} \times \mathcal{Y}_{v})$. Its purpose is to define the relationship of an implementation belonging to a platform.
\end{MyDef}

\begin{MyDef}[Containment relationship of ] Given a composite module implementation $\mathcal{X}_{u}$ and a simple module implementation $\mathcal{Y}_{v}$ or another implementation of a composite module $\mathcal{X}_{u'}$, the relationship of containment $\mathcal{R}_{\mathcal{C}}$ is defined as: $\mathcal{R}_{\mathcal{C}} \subseteq (\mathcal{X}_{u} \times \mathcal{Y}_{v}) \vee (\mathcal{X}_{u} \times \mathcal{X}_{u'})$. Its purpose is to define the relationship of containment between two module implementations.
\end{MyDef}

\begin{MyDef}[Configuration] A configuration $\mathcal{F}$ is determined by a finite set $\mathcal{F} = \{\mathcal{Z}\}$, where: $\mathcal{Z}$ is the set of statements with the implementation relationships $(\mathcal{Z} \neq \varnothing)$.
\end{MyDef}

\begin{MyDef}[Implementation relationship] Given a module $\mathcal{M}_{n}$ and the implementation of this module $\mathcal{W}_{n}$, the relationship of implementation $\mathcal{R}_{\mathcal{I}}$ is defined as $\mathcal{R}_{\mathcal{I}} \subseteq \mathcal{M}_{n} \times \mathcal{W}_{n}$. Its purpose consists of defining the relationship between a module and its corresponding implementation. 
\end{MyDef}

The framework defined above is difficult to understand without explaining its behavior when it is applied to a domain or example scenario. For this reason, the following section is intended to illustrate a case study of an information systems which has been successfully designed and developed based our proposal.

\section{Case Study: SOLERES-KRS}
\label{sec:casestudy}

\begin{figure}[!b]
\begin{center}
\includegraphics[width=8.301cm]{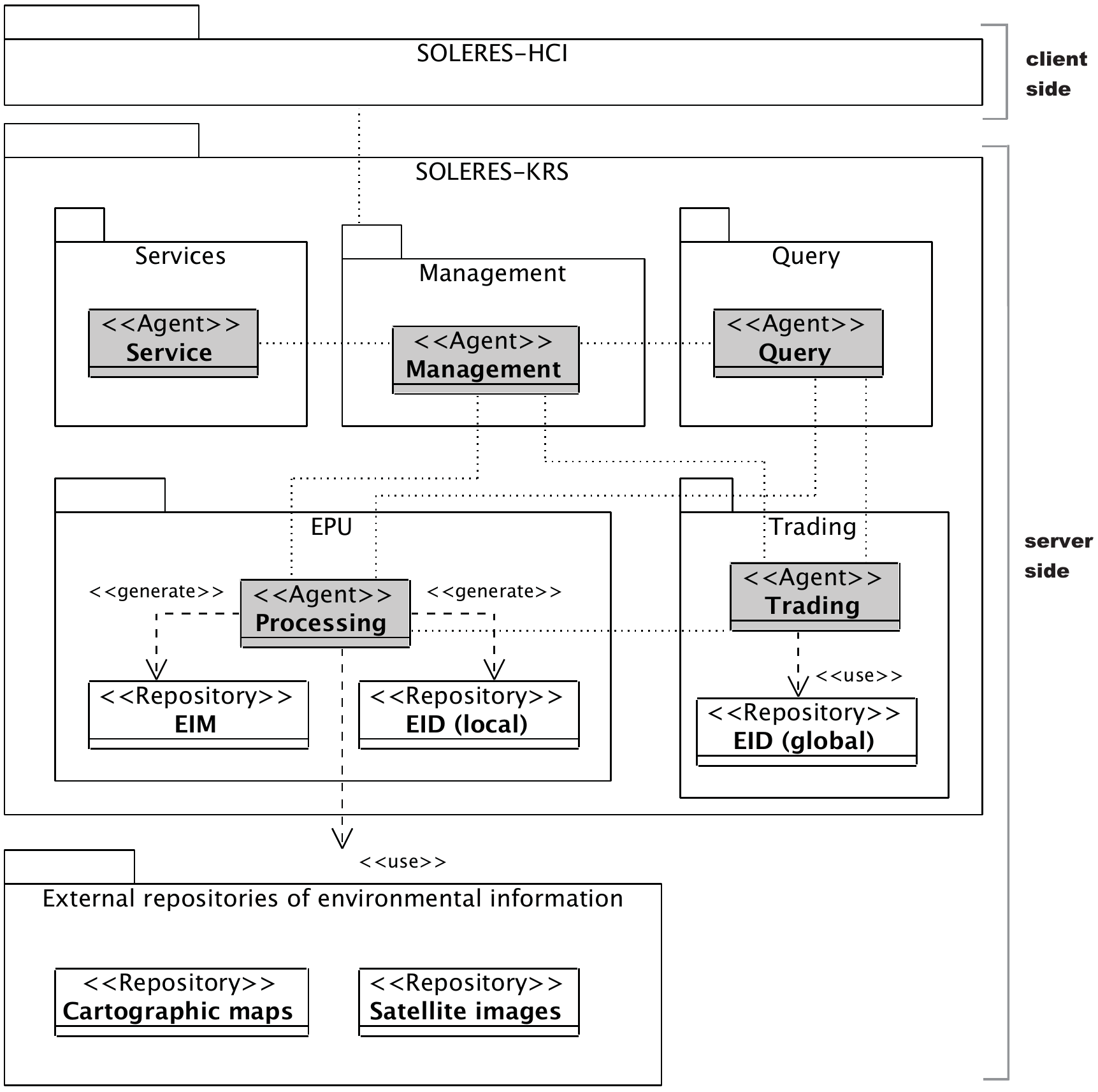}
\end{center}\caption{{\color{black}SOLERES system Architecture}}
\label{f12}
\end{figure}

This section presents a case study for application of the TKRS framework in Environmental Management Information Systems (EMIS), in which large volumes of environmental data are usually used. Some environmental authorities, organizations or institutions use ecological maps for their evaluation of environmental impact, territorial management, monitoring impact of certain activities on protected spaces, etc. The cost of making these ecological maps is very high, because they are based on fieldwork and periodic updating (yearly). However, in some environmental management actions, where studies of the current situation in the territory are needed (for example, evaluation of growth of greenhouses in eastern Almer\'ia Province), these ecological maps do not provide the updated environmental information required for decision-making, and new fieldwork is necessary. This environmental information is not available in other spatial information sources, such as satellite (or flight) images, which is much faster (daily) and more economical to acquire.

Prior studies have explored the possibility and feasibility of establishing a correlation between environmental variables (cartography) used in making ecological maps and the information from satellite images (correlation finally demonstrated) that would enable them to be acquired automatically, much faster and at a much lower economic and human cost. To do this, an ecological map is acquired based on a map or a set of cartographic maps of the study area, and by sectorization (which receives another further series of inputs). Then, based on a satellite image or set of images of the same zone, by classification (which also receives another series of inputs), a classified image is acquired. Once the ecological map has been made and the image has been classified, a neural network correlates them. 

All the information used during the above processes must somehow be stored and managed. Therefore, another study consisted of developing an information management and usage system (in this case environmental) to assist human-computer interaction (HCI) with user interfaces that adapt to user profile habits \cite{r30, r31, r16}, and with intelligent software agents that mediate for users in information search and usage.
{\color{black}Thus, decision-making and prediction/prevention tasks (which are addressed in the majority of the EMIS and GIS) are facilitated.}

This system, called SOLERES, can be observed in Figure \ref{f12}. It is in turn made up of two subsystems, (1) the Knowledge Representation System (SOLERES-KRS or SKRS), responsible for environmental information management, and (2) SOLERES-HCI \cite{r28, r4}, related to the user interfaces that assist in making use of this information. This section concentrates exclusively on the development of the SOLERES-KRS following the design of TKRS system framework. The information domain which the SOLERES-KRS manages is basically two large, bulky sources of original information: (a) cartographic information, and (b) information from satellite images of the study area (regions in southeast Spain). Given the volume of information, and that it can come from different sources, a distributed solution was chosen in which the information is distributed into units or areas called Environmental Process Units (EPU in Figure \ref{f12}), which work with the metadata found from it, \textit{i.e.}, the Environmental Information Maps (EIM). The Trading modules have a dual function: (i) integrating the information from the different EPUs and (ii) facilitating its search and retrieval. As mentioned in Section \ref{sec:modeling}, trading modules use new meta-metadata based on from the EIMs, which are stored locally in a repository (also generated by the EPUs, and therefore, a copy is stored in another local repository). These ``meta-metadata'' are called Environmental Information metaData (EID). The modules mentioned, along with Query, Management and Services, which appear in Figure \ref{f12}, make up the TKRS architecture. {\color{black} EID and EIM documents form the specific kind of information managed by the TKRS framework in this case study. Therefore, the implementation of the operations linked to the database storing the information must be modified according to its structure if the TKRS framework is applied to other domains.}

Finally, it should be emphasized that each system user would receive support from a SOLERES-HCI agent which would manage presentation and interaction with the user interface, mediating between the user it represents and other users in the system by means of their respective agents, and also managing environmental queries, like a virtual consultant or advisor who interacts with other SOLERES-KRS subsystem agents.

\subsection{Design of the system architecture}
\label{s61}

\begin{figure}[!b]
\begin{center}
\includegraphics[width=\textwidth]{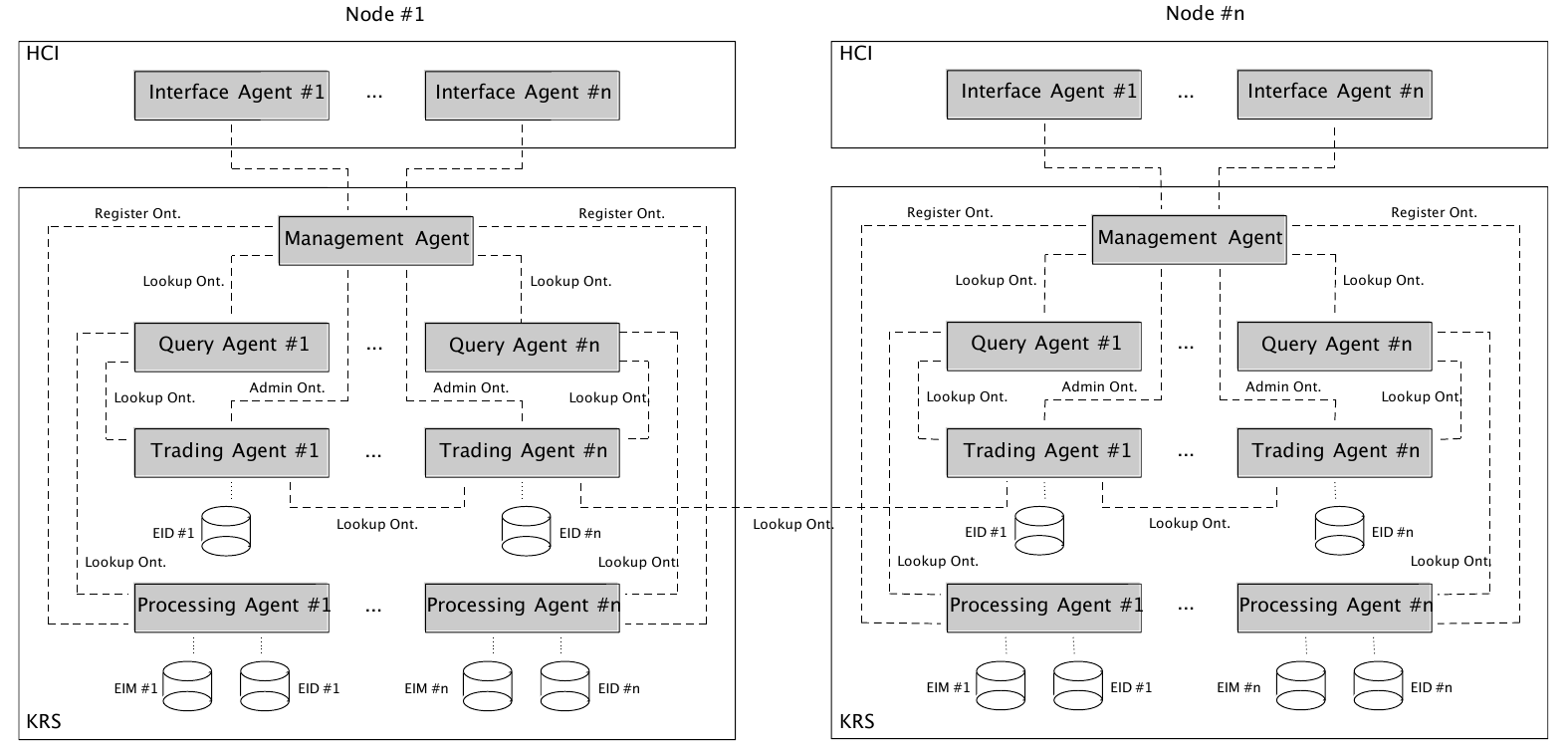}
\end{center}\caption{SOLERES-KRS general architecture}
\label{f13}
\end{figure}

Briefly, the TKRS architecture described in Section \ref{sec:framework} is made up of a set of nodes, each of which consists in turn of a series of modules: services, management, query, trading and processing modules. These modules are required or not by the nodes. Keeping this architecture model in mind, the choice of agent technology for implementation of the SOLERES-KRS and its characteristics as described, the general system architecture would be defined as shown in Figure \ref{f13} \cite{r29, r17} (this diagram does not include agents related to the service module). 

The management module is implemented with a Management Agent, which acts as a mediator between the SOLERES-HCI Interface Agents and the rest of the SOLERES-KRS agents to manage all the user demands. The query module, with a Query Agent, solves the ecological information queries directly, the trading module facilitates search and location of this information using the EIDs stored in its repository using a Trading Agent, and the processing module, with a Processing Agent, which represents the EPUs, is responsible for the knowledge sources (management of EIMs, export of EIDs to the associated trading agent, etc.). There are a series of agents in the service module typical of the development environment used (JADE), such as the Agent Management System (AMS) or the Directory Facilitator (DF), which provides a white pages service for managing all the agents, and another yellow pages service.

\begin{figure}[!b]
	\begin{center}
		\includegraphics[width=1.00\textwidth]{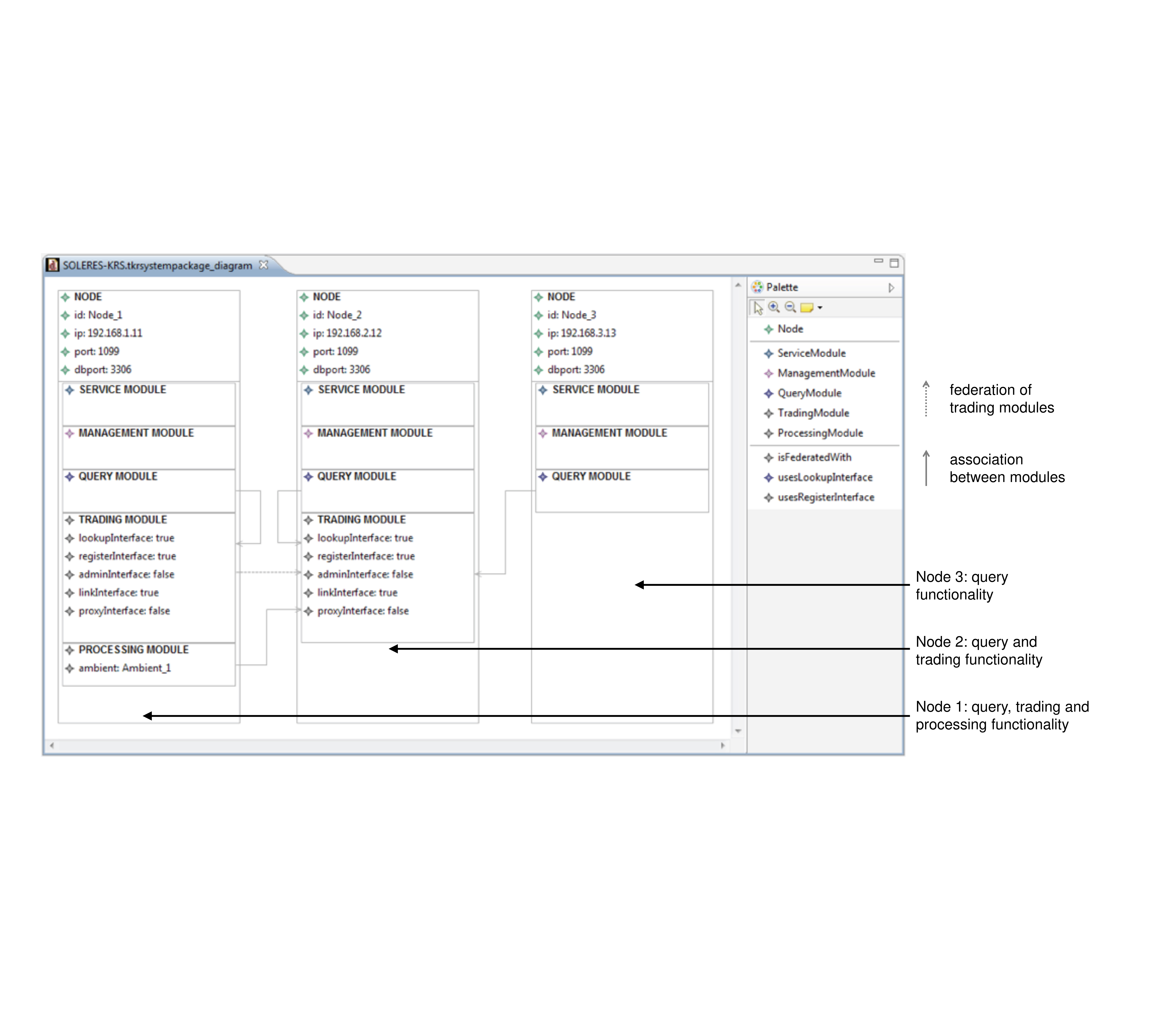}
	\end{center}\caption{{\color{black}SOLERES-KRS architecture model made up of three nodes and represented using the GMF tool}}
	\label{f14}
\end{figure}

Figure \ref{f14} does show the concrete SOLERES-KRS architecture model that was {\color{black}constructed using the TKRS architecturel metamodel of Section 2. This model has been designed using the GMF editor mentioned above, which allows us to build models conforming the metamodel shown in Figure \ref{f2}}. As observed, the system is made up of three nodes, each of which has a management agent (which make up the TKRS management module), a query agent (query module) and agents corresponding to the obligatory service module. {\color{black}``Node 1'' and ``Node 2' also contain a trading agent (trading module) and, additionally, ``Node 1'' is made up of a processing agent (processing module, EPU), meeting the minimum TKRS demands.} The trading agents in the above nodes are federated (shown with a dashed arrow) and their query agents are associated with the trading agents in their respective nodes, while the processing agent in ``Node 1'' and the query agent in ``Node 3'' are associated with the trading agent in ``Node 2'' (all of these associations are shown with continuous arrows). 

\subsection{Implementation based on a Multi-Agent System}
\label{s62}

\begin{figure}
\begin{center}
\includegraphics[width=9.8cm]{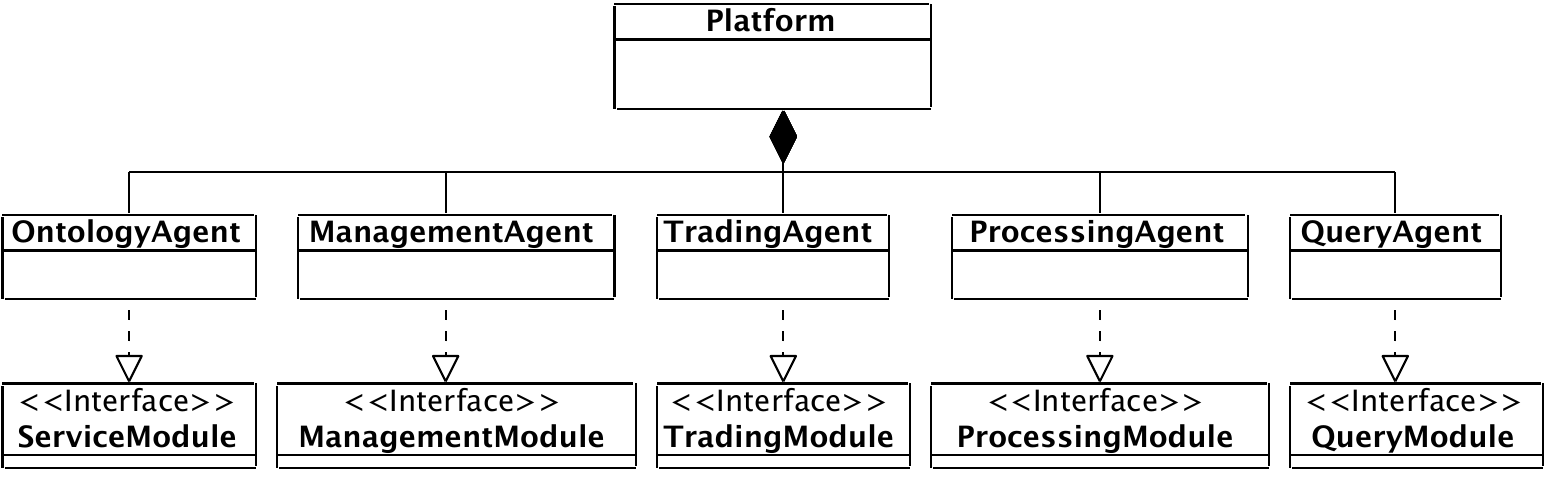}
\end{center}\caption{Conceptual model of a TKRS implementation based on MAS}
\label{f15}
\end{figure}

\begin{figure}
\begin{center}
\includegraphics[width=7.9cm]{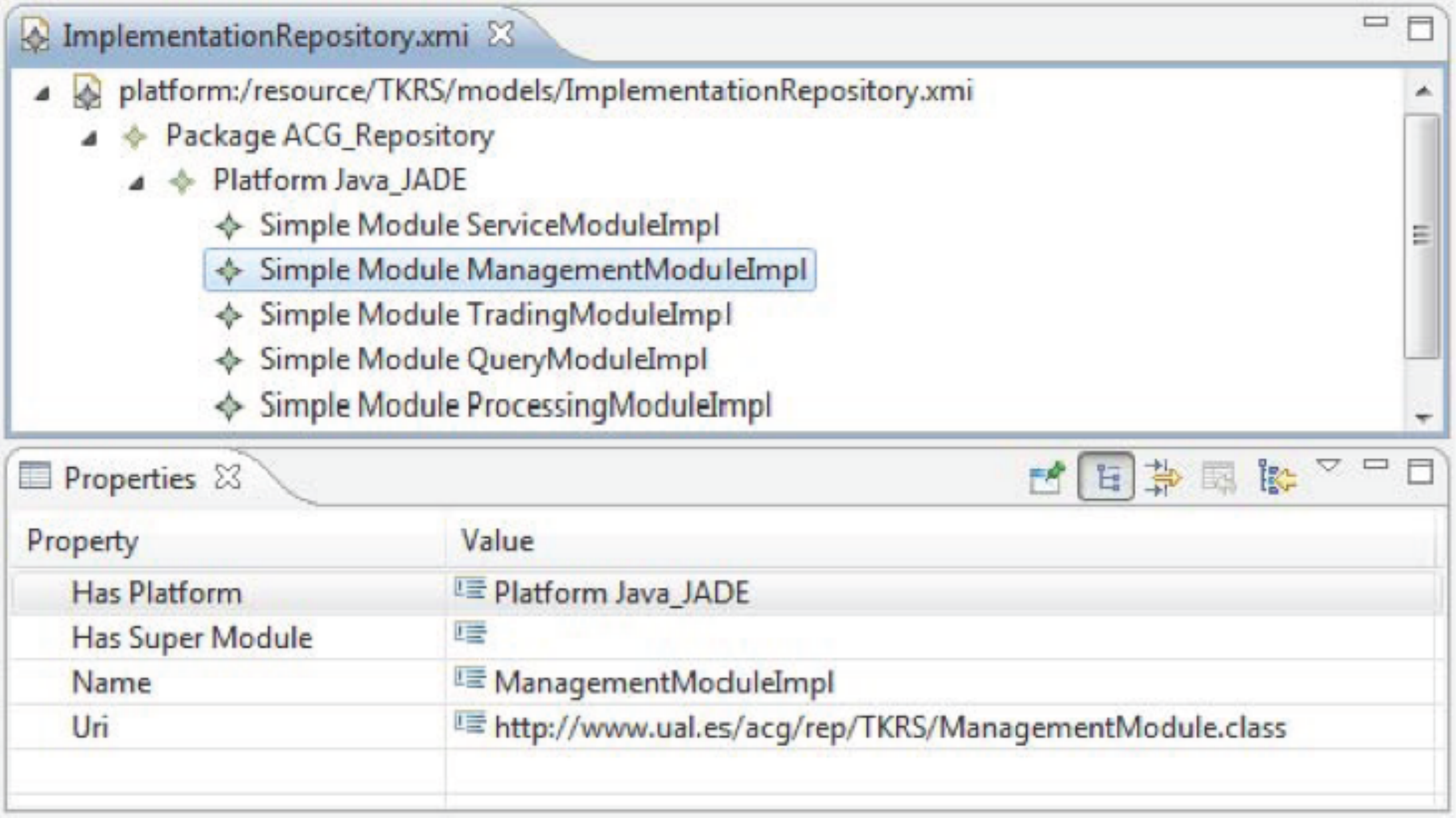}
\end{center}\caption{Repository model using Eclipse Reflective Ecore Model Editor}
\label{f16}
\end{figure}

The selection of the Multi-Agent System (MAS) technology for implementing the TKRS was determined by the characteristics of software agents (autonomy, persistence, sociability, reactivity, pro-activity, intelligence, etc.). Basically, as shown in the section above and the conceptual model in Figure \ref{f15}, each of the system modules can be implemented by an agent or set of agents. However, in the service module, a specialized agent can be created for each type of service. For example, there may be an ontology agent for managing ontologies, so when any system agent needs to send a message to another agent making use of one of the service/process ontologies or needs to manage a metadata or meta-metadata register, to avoid having to specialize in these tasks, it uses an ontology agent. Figure \ref{f16} shows the model of a repository (PIM/M1) with TKRS implementation based on MAS as described for a Java environment, using the reflective ecore model editor. The top of this figure shows how each of the system modules is implemented as a simple module, while the bottom shows management module implementation properties: their name, location and the platform developed in. 

It was already advanced in Section \ref{sec:casestudy}, that SOLERES-KRS was going to be made up of a set of distributed EPUs (working independently in system knowledge management), the Environmental Information Maps (EIM) and their corresponding Environmental Information metaData (EID). Cooperation among them was achieved by developing a trading service based on the OntoTrader specification and implemented with trading agents (or \textit{traders}). These agents act as mediators (although indirectly) between the SOLERES-HCI interface agents and the EPU processing agents (\textit{exporters}) to satisfy user information demands. For each trading agent to be able to perform its information search task, each EPU exports its EIDs to the trading agent with which it is associated, so it has an overall repository in which the EIDs of all the EPUs associated with it are stored. The query agents have an \textit{importer} role, and their function is to find the information demanded by the users by means of the trading agent, either with the meta-metadata it stores or else with metadata filtered by the trading agent and stored by the processing agents. The management agent is the trading service \textit{administrator}, which defines, administers and verifies that the service rules are complied with. This agent is also the direct mediator with SOLERES-HCI and controls the rest of the subsystem environment by planning the tasks performed. 

The three service/process ontologies specified in Section \ref{sec:framework} were implemented to define the protocols for behavior and interaction with trading agents through their interfaces, and by extension, with any other system agent \cite{r6}. When one agent needs another to perform a specific action, it constructs a message in which the action is described using one of these ontologies and sends it to the second agent. This one receives the message, extracts the ontology content, performs the appropriate action and uses the same ontology to show the result of the action to the first. This process is repeated for the various actions in which more than one agent intervene. Table \ref{t7} gives the definition of the Lookup ontology in OWL/XML notation, where there are three main classes, \textit{Concept}, \textit{Action} and \textit{Predicate}, from which the rest inherit. For example, \textit{QueryForm} is a subclass of \textit{Concept}, \textit{Query} of \textit{Action} and \textit{QuerryError} of \textit{Predicate}. The type, domain and range are defined for each of the concept properties. Thus the domain of the \textit{queryFormURI} property is the \textit{DatatypeProperty} type and its range is the data string type. 

\begin{table}
\centering
\caption{Definition of the Lookup ontology in OWL/XML}
\label{t7}
{\fontsize{6pt}{7pt}\selectfont
\begin{tabular}{p{13cm}}
\hline
\vspace{-2mm}
\begin{verbatim}
01  <?xml version="1.0"?>
02
03  <!DOCTYPE rdf:RDF [
04    <!ENTITY lookup "http://www.ual.es/acg/ont/TKRS/LookupOntology.owl#">
05    <!ENTITY owl "http://www.w3.org/2002/07/owl#">
06    <!ENTITY rdf "http://www.w3.org/1999/02/22-rdf-syntax-ns#">
07    <!ENTITY rdfs "http://www.w3.org/2000/01/rdf-schema#">
08    <!ENTITY xsd "http://www.w3.org/2001/XMLSchema#"> ]>
09
10  <rdf:RDF xml:base = "&lookup;" xmlns = "&lookup;" xmlns:lookup = "&lookup;" 
11     xmlns:owl = "&owl;" xmlns:rdf = "&rdf;" xmlns:rdfs = "&rdfs;" xmlns:xsd = "&xsd;">
12    <owl:Ontology rdf:about="&lookup;"><rdfs:label>Lookup Ontology</rdfs:label></owl:Ontology>
13    <owl:Class rdf:ID="Concept"/>
14    <owl:Class rdf:ID="Action"/>
15    <owl:Class rdf:ID="Predicate"/>
16    <owl:Class rdf:ID="QueryForm"><rdfs:subClassOf rdf:resource="#Concept"/></owl:Class>
17    <owl:FunctionalProperty rdf:ID="queryFormURI">
18      <rdf:type rdf:resource="&owl;DatatypeProperty"/>
19      <rdfs:domain rdf:resource="#QueryForm"/>
20      <rdfs:range rdf:resource="&xsd;string"/>
21    </owl:FunctionalProperty>
22    <owl:Class rdf:ID="OfferSeq"><rdfs:subClassOf rdf:resource="#Concept"/></owl:Class>
23    <owl:FunctionalProperty rdf:ID="offerSeqURI">
24      <rdf:type rdf:resource="&owl;DatatypeProperty"/>
25      <rdfs:domain rdf:resource="#OfferSeq"/>
26      <rdfs:range rdf:resource="&xsd;string"/>
27    </owl:FunctionalProperty>
28    <owl:Class rdf:about="#Query"><rdfs:subClassOf rdf:resource="#Action"/></owl:Class>
30    <owl:Class rdf:ID="QueryError"><rdfs:subClassOf rdf:resource="#Predicate"/></owl:Class>
31    <owl:Class rdf:ID="EmptyOfferSeq"><rdfs:subClassOf rdf:resource="#Predicate"/></owl:Class>
32    <owl:Class rdf:ID="NotEmptyOfferSeq"><rdfs:subClassOf rdf:resource="#Predicate"/></owl:Class>
33    ...
34  </rdf:RDF>

\end{verbatim}
\\
\hline
\end{tabular}
}
\end{table}

\subsection{System configuration}
\label{s63}

Briefly reviewing the TKRS framework \cite{r5}, when the system architecture and the implementation repository model have been defined using the system configuration model and a series of M2M and M2T transformations, it can be deployed in a given environment. This is precisely what has been done below. The SOLERES-KRS architecture model and the agent-based TKRS implementation repository model were defined in the two sections above, respectively. For their elements to be referenced by the configuration model, each was transformed by a M2M operation, resulting in the new representation which appears in Tables \ref{t8} and \ref{t9}. The generated models are based on the grammar developed in Section \ref{sec:modeling}. 

In Table \ref{t8}, which shows the architecture model, note that the TKRS entity is followed by the system name, SOLERES-KRS, and contains three Node entities corresponding to its three nodes (lines \#3 to \#28, \#29 to \#48 and \#49 to \#60, respectively). Each of them is followed by the name of the node (for example, the first, \textit{Node\_1}) and in parentheses the attributes (\textit{ip}, \textit{port}, etc.) followed by their value (in the case of the ip address of \textit{Node\_1}, ``192.168.1.11''). Then, modules include their type (\textit{ServiceModule}, \textit{ManagementModule}, etc.) followed by their name (\textit{ServiceModule\_1\_1}, \textit{ManagementModule\_1\_1}, etc.). The corresponding attributes are also defined in parentheses (those modules which have them), and the node it belongs to is given by the relationship \textit{hasNode}. Finally, each node refers to the system in which it operates through the relationship \textit{hasTKRS}. Similarly, Table \ref{t9} shows the repository model, where for the \textit{Java\_JADE} platform, five \textit{SimpleModules} are specified (lines \#5 to \#24), each followed by its name (for example, the first, \textit{ServiceModuleImpl}) and they define the attribute \textit{uri} (``http//.../acg/rep/TKRS/ServiceModule.class''), as well as the platform to which it belongs (``Java\_JADE'') through the relationship \textit{hasPlatform}. 

\begin{table}
\centering
\caption{SOLERES-KRS architecture model expressed using the grammar defined}
\label{t8}
{\fontsize{6pt}{7pt}\selectfont
\begin{tabular}{p{9cm}}
\hline
\vspace{-3mm}
\begin{verbatim}
01  Package SOLERES
02  TKRS SOLERES_KRS {
03    Node Node_1 {
04      ip "192.168.1.11"
05      port "1099"
06      dbport "3306"
07      ServiceModule ServiceModule_1_1 { hasNode Node_1 }
08      ManagementModule ManagementModule_1_1 { hasNode Node_1 }
09      QueryModule QueryModule_1_1 {
10        usesLookupInterface TradingModule_1_1
11        hasNode Node_1
12      }
13      TradingModule TradingModule_1_1 {
14        usesLookupInterface true
15        usesRegisterInterface true
16        usesAdminInterface false
17        usesLinkInterface true
18        usesProxyInterface false
19        isFederatedWith "Node_2.TradingModule_2_1"
20        hasNode Node_1
21      }
22      ProcessingModule ProcessingModule_1_1 {
23        ambient Ambient_1
24        usesRegisterInterface "Node_2.TradingModule_2_1"
25        hasNode Node_1
26      }
27      hasTKRS SOLERES_KRS
28    }
29    Node Node_2 {
30      ip "192.168.1.12"
31      port "1099"
32      dbport "3306"
33      ServiceModule ServiceModule_2_1 { hasNode Node_2 }
34      ManagementModule ManagementModule_2_1 { hasNode Node_2 }
35      QueryModule QueryModule_2_1 {
36        usesLookupInterface TradingModule_2_1
37        hasNode Node_2
38      }
39      TradingModule TradingModule_2_1 {
40        usesLookupInterface true
41        usesRegisterInterface true
42        usesAdminInterface false
43        usesLinkInterface true
44        usesProxyInterface false
45        hasNode Node_2
46      }
47      hasTKRS SOLERES_KRS
48    }
49    Node Node_3 {
50      ip "192.168.1.13"
51      port "1099"
52      dbport "3306"
53      ServiceModule ServiceModule_3_1 { hasNode Node_3 }
54      ManagementModule ManagementModule_3_1 { hasNode Node_3 }
55      QueryModule QueryModule_3_1 {
56        usesLookupInterface "Node_2.TradingModule_2_1"
57        hasNode Node_3
58      }
59      hasTKRS SOLERES_KRS
60    }
61  }

\end{verbatim}
\\
\hline
\end{tabular}
}
\end{table}

\begin{table}
\centering
\caption{TKRS repository model expressed using the grammar defined}
\label{t9}
{\fontsize{6pt}{7pt}\selectfont
\begin{tabular}{p{9cm}}
\hline
\vspace{-2mm}
\begin{verbatim}
01  Package UAL_Repository
02
03  ImplementationRepository {
04    Platform Java_JADE {
05      SimpleModule ServiceModuleImpl {
06        uri "http://.../acg/rep/TKRS/ServiceModule.class"
07        hasPlatform Java_JADE
08      }
09      SimpleModule ManagementModuleImpl {
10        uri "http://.../acg/rep/TKRS/ManagementModule.class"
11        hasPlatform Java_JADE
12      }
13      SimpleModule QueryModuleImpl {
14        uri "http://.../acg/rep/TKRS/QueryModule.class"
15        hasPlatform Java_JADE
16      }
17      SimpleModule TradingModuleImpl {
18        uri "http://.../acg/rep/TKRS/TradingModule.class"
19        hasPlatform Java_JADE
20      }
21      SimpleModule ProcessingModuleImpl {
22        uri "http://.../acg/rep/TKRS/ProcessingModule.class"
23        hasPlatform Java_JADE
24      }
25    }
26  }

\end{verbatim}
\\
\hline
\end{tabular}
}
\end{table}

The SOLERES-KRS configuration model can now finally be defined (Table \ref{t10}), where by means of a set of sentences (\textit{Statements}), the models that form part of each system node are associated with their corresponding implementation in the repository (following the PIM/PSM architecture defined). For example, the first of them (lines \#4 to \#6) is associated with the \textit{SOLERES.KRS.Node\_1.ServiceModule\_1\_1} in the architecture by means of the \textit{hasTKRSModule} relationship with the \textit{ACG\_Repository.Java\_JADE.ServiceModuleImpl} implementation (by the relationship \textit{hasImplementationRepositoryModule}) and so on. 

\begin{table}
\centering
\caption{SOLERES-KRS configuration model expressed using the grammar defined}
\label{t10}
{\fontsize{6pt}{7pt}\selectfont
\begin{tabular}{p{12cm}}
\hline
\vspace{-2mm}
\begin{verbatim}
01  Package SOLERES_Configuration
02  
03  Configuration {
04    Statement {
05      hasTKRSModule "SOLERES.KRS.Node_1.ServiceModule_1_1"
06      hasImplementationRepositoryModule "ACG_Repository.Java_JADE.ServiceModuleImpl" }
07    Statement {
08      hasTKRSModule "SOLERES.KRS.Node_1.ManagementModule_1_1"
09      hasImplementationRepositoryModule "ACG_Repository.Java_JADE.ManagementModuleImpl" }
10    Statement {
11      hasTKRSModule "SOLERES.KRS.Node_1.QueryModule_1_1"
12      hasImplementationRepositoryModule "ACG_Repository.Java_JADE.QueryModuleImpl" }
13    Statement {
14      hasTKRSModule "SOLERES.KRS.Node_1.TradingModule_1_1"
15      hasImplementationRepositoryModule "ACG_Repository.Java_JADE.TradingModuleImpl" }
16    Statement {
17      hasTKRSModule "SOLERES.KRS.Node_1.ProcessingModule_1_1"
18      hasImplementationRepositoryModule "ACG_Repository.Java_JADE.ProcessingModuleImpl" }
19    Statement {
20      hasTKRSModule "SOLERES.KRS.Node_2.ServiceModule_2_1"
21      hasImplementationRepositoryModule "ACG_Repository.Java_JADE.ServiceModuleImpl" }
22    Statement {
23      hasTKRSModule "SOLERES.KRS.Node_2.ManagementModule_2_1"
24      hasImplementationRepositoryModule "ACG_Repository.Java_JADE.ManagementModuleImpl" }
25    Statement {
26      hasTKRSModule "SOLERES.KRS.Node_2.QueryModule_2_1"
27      hasImplementationRepositoryModule "ACG_Repository.Java_JADE.QueryModuleImpl" }
28    Statement {
29      hasTKRSModule "SOLERES.KRS.Node_2.TradingModule_2_1"
30      hasImplementationRepositoryModule "ACG_Repository.Java_JADE.TradingModuleImpl" }
31    Statement {
32      hasTKRSModule "SOLERES.KRS.Node_3.ServiceModule_3_1"
33      hasImplementationRepositoryModule "ACG_Repository.Java_JADE.ServiceModuleImpl" }
34    Statement {
35      hasTKRSModule "SOLERES.KRS.Node_3.ManagementModule_3_1"
36      hasImplementationRepositoryModule "ACG_Repository.Java_JADE.ManagementModuleImpl" }
37    Statement {
38      hasTKRSModule "SOLERES.KRS.Node_3.QueryModule_3_1"
39      hasImplementationRepositoryModule "ACG_Repository.Java_JADE.QueryModuleImpl" }
40  }

\end{verbatim}
\\
\hline
\end{tabular}
}
\end{table}

Finally, the code for deploying the system in a real environment is obtained based on the definition of this configuration model and the M2T transformation, presented in Subsection \ref{sec:modeling}. This transformation generates several files: a script file (see Table \ref{t11}) with operating system commands such as \texttt{mkdir}, \texttt{cd} and \texttt{wget} to create a folder structure that includes the system implementation files, and for compiling (\texttt{javac TKRS.java}), and there is a set of Java files (one for each node in the system) located in the corresponding structure folder. In this files, configuration and initialization operations (in the class constructor) of all its parameters (class attributes) are performed based on the specific values in the architecture design. 

\begin{table}
\centering
\caption{Script file for SOLERES-KRS deployment}
\label{t11}
{\fontsize{6pt}{7pt}\selectfont
\begin{tabular}{p{9cm}}
\hline
\begin{verbatim}
01  #!/bin/bash
02  clear
03  mkdir /SOLERES_KRS/Node_1/modules
04  cd /SOLERES_KRS/Node_1/modules
05  wget http://.../acg/rep/TKRS/ServiceModule.class
06  wget http://.../acg/rep/TKRS/ManagementModule.class
07  wget http://.../acg/rep/TKRS/TradingModule.class
08  wget http://.../acg/rep/TKRS/QueryModule.class
09  wget http://.../acg/rep/TKRS/ProcessingModule.class
10  cd /SOLERES_KRS/Node_1
11  javac TKRS.java
12  mkdir /SOLERES_KRS/Node_2/modules
13  cd /SOLERES_KRS/Node_2/modules
14  wget http://.../acg/rep/TKRS/ServiceModule.class
15  wget http://.../acg/rep/TKRS/ManagementModule.class
16  wget http://.../acg/rep/TKRS/TradingModule.class
17  wget http://.../acg/rep/TKRS/QueryModule.class
18  cd /SOLERES_KRS/Node_2
19  javac TKRS.java
20  mkdir /SOLERES_KRS/Node_3/modules
21  cd /SOLERES_KRS/Node_3/modules
22  wget http://.../acg/rep/TKRS/ServiceModule.class
23  wget http://.../acg/rep/TKRS/ManagementModule.class
24  wget http://.../acg/rep/TKRS/QueryModule.class
25  cd /SOLERES_KRS/Node_3
26  javac TKRS.java
27  cd /

\end{verbatim}
\\
\hline
\end{tabular}
}
\end{table}

\begin{figure}
	\centering
	\hspace{-4cm}
	\begin{minipage}[l]{5cm}
		\includegraphics[width=4cm]{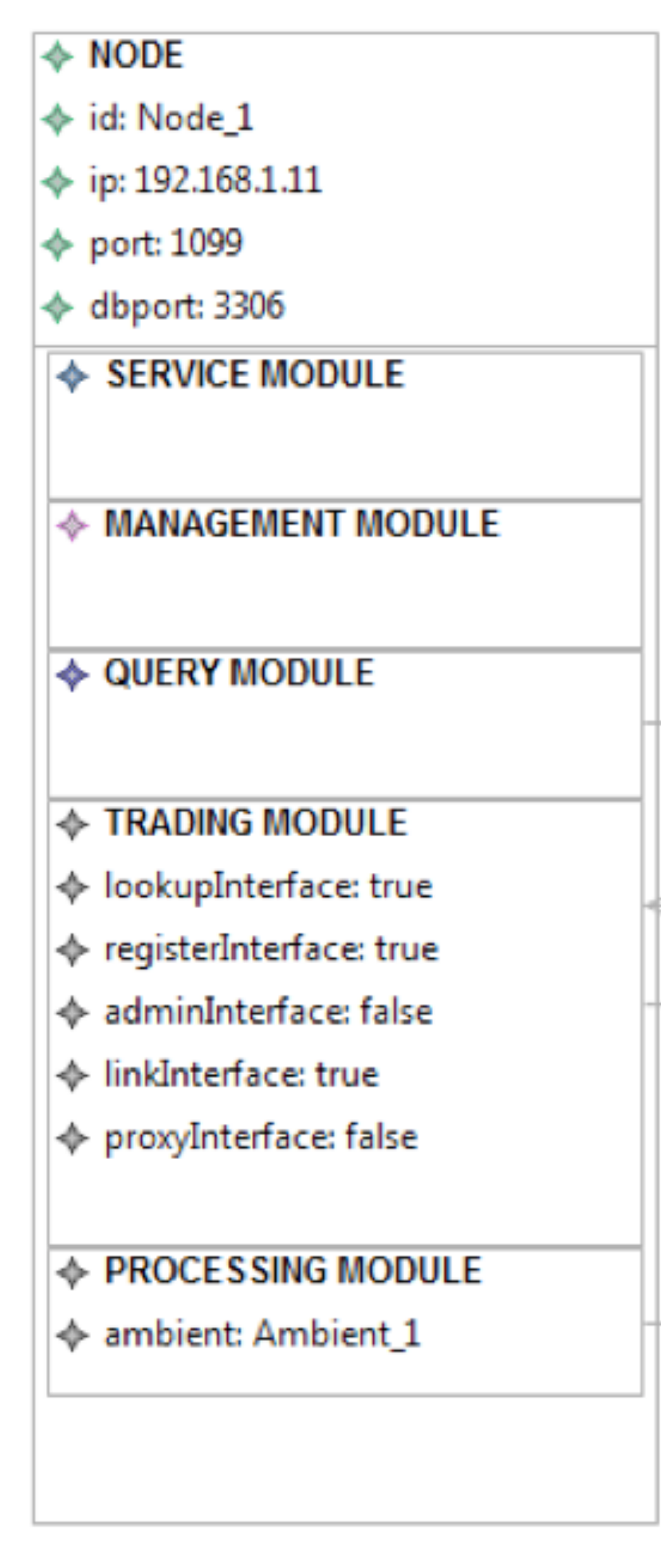}
	\end{minipage}
	\begin{minipage}[r]{4cm}
		\delimitershortfall=0pt
		\setlength{\dashlinegap}{1.5pt}
		{\fontsize{6pt}{7pt}\selectfont
			\begin{tabular*}{1pt}{: p{1cm}}
				\begin{verbatim}
				1  package SOLERES_KRS.Node_1;
				2
				3  import SOLERES_KRS.Node_1.modules.*;
				4
				5  public class TKRS {
				6    private String ip = null;
				7    private int port = -1;
				8    private int dbport = -1;
				9    private ServiceModule serviceModule = null;
				10    private ManagementModule managementModule = null;
				11    private QueryModule queryModule = null;
				12    private TradingModule tradingModule = null;
				13    private ProcessingModule processingModule = null;
				14
				15    public TKRS() {
				16      this.ip = "192.168.1.11";
				17      this.port = 1099;
				18      this.dbport = 3306;
				19      this.serviceModule = new ServiceModule();
				20      this.managementModule = new ManagementModule();
				21      this.queryModule = new QueryModule();
				22      this.tradingModule = new TradingModule();
				23      this.processingModule = new ProcessingModule();
				24      // CODE
				25    }
				26    // CODE
				27  }
				
				
				\end{verbatim}
				\\
			\end{tabular*}
		}\end{minipage}
		\caption{{\color{black}Java file for configuration of SOLERES-KRS Node 1}}\label{ffinal}
	\end{figure}

Figure \ref{ffinal} shows the content of one of these Java files, which corresponds to the implementation of the corresponding node defined in Figure \ref{f14} (extracted from it and placed next to this table for the explanation). It shows the TKRS class specification, which forms part of the \textit{SOLERES\_KRS.Node\_1} (line \#1) package, where Node 1 of the SOLERES-KRS is configured. The classes that implement the different TKRS modules are imported in line \#3. The attributes are stated in lines \#6-\#13 and among them are both node properties (\textit{ip}, \textit{port}, etc.), and the TKRS modules it includes (in this case, all). In the class constructor (lines \#15-\#25), the above attributes are initialized with the values shown by means of the graphics tool, etc. This is repeated for Nodes 2 and 3 of the system architecture. 

\section{Related work}
\label{sec:rw}

As mentioned above, one of the biggest challenges of the Information Society consists of addressing the continuous growth of the information volume handled in most areas, but not only its collection or storage, but also its processing and analysis. The concept of Big Data arises around this kind of operations. In \cite{madden2012from}, the author writes about its origin and presents its relation with the data processing platforms. Meanwhile, in \cite{kaisler2014introduction}, a review is performed presenting the four main challenges (known as the four Vs) of Gartner (increasing Volume of data, increasing Velocity, increasing Variety of data types and structures, and increasing Variability of data), and the authors propose the fifth V of Big Data, the Value, in order to improve decision-making operations. \cite{sheth2014deriving} focuses on the challenge of transforming big data into smart data by using metadata (in the form of ontologies, domain models, or vocabularies), and employing semantics and intelligent processing mechanisms. \cite{ishwarappa2015brief} also reviews the five Vs of Big Data, as well as the technologies used for its management. Frequently, the information managed is unstructured and does not follow any model; in contrast, we propose an approach based on MDE techniques thus facilitating structured representation and management. \cite{bakshi2012} focuses on the analysis of this type of information and shows the Apache Hadoop Project and MapReduce as an approximation. In \cite{kune2016anatomy} the authors also discuss the evolution of the Big Data computing, its taxonomy, the technologies, etc., but they progress and integrate it with the concept of cloud. The proposal presented in this paper is based on an Ontological Web-Trading Model, as another approach to support the Vs of the Big Data mentioned above.

There are not really many studies related to the framework for design and development of Web-based Information Management Systems that use trading and ontologies as the solution for handling Big Data. The characteristics of most of the trading services in the literature \cite{r7, r44, r34, r8, r39, r46, r48} are similar, because they are all based on the ODP \cite{r32} trading model (also standardized by the OMG). One of the studies in this area that seems to differ from the rest is WebTrader \cite{r54}, a study that presents a trading model for developing Internet-based software and uses XML to describe component services. The Pilarcos Project, a research project of the TEKES National Technology Agency of Finland, Helsinki University Computer Science Department and Nokia, SysOpen and Tellabs should also be mentioned. It defends, among other characteristics, the need for a trading service based on ODP/OMG for developing middleware solutions with automatic management of federated applications. The main drawback of these proposals is the use of only component specifications. 

There are also ODP trading service implementation approaches that differ in their interaction and communication among objects or components, or description of the communication language. For example, in \cite{r15}, the authors present a trading service called DOKTrader, which acts on a federated database system, the Distributed Object Kernel (DOK). Another example may be found in \cite{r38}, in which a framework for developing distributed applications in a Common Open Service Market (COSM) is created, making use of a Service Interface Description Language (SIDL) to describe the services managed by the trader. \cite{r35} present a web market service driven by ontologies. In this system, an ontological communication language is used for representing queries, offers and agreements. In \cite{r53}, an ontology is also used to describe and facilitate Dynamic Process Collaboration (DPC). The OntoTrader model which our TKRS framework uses, attempts to provide a trading service promoting both interoperability of objects or web components in open distributed systems, and improve search and retrieval of the information in them. This model not only enables information management in any application domain, but also the specifications of the services offered by the components, and sets a common standard for communication with the trading service through its various interfaces. To do this it makes use of a series of ontologies, more specifically, a data ontology and a service/process ontology for each of these interfaces. 

There are many independent examples of ontology application in systems where information semantics are fundamental. For example, in \cite{r12}, the authors present an environmental decision support system called OntoWEDSS, where an ontology is used to model wastewater treatment, providing a common vocabulary and explicit conceptualization. Another example may be found in \cite{r19}, where an air quality monitoring system uses an ontology to define messages and communication actions concisely and without ambiguities. In \cite{r11}, the authors present Ecolingua, an EngMath family ontology, for representing quantitative ecological data. All these examples show the use of ontologies for constructing models that describe the entities in a given domain and to characterize their associated relationships and restrictions. In \cite{r60}, the authors present an ontology for representing geographic data and related functions. To satisfy the need for an interoperable GIS, in \cite{r3}, the authors propose the design of a geo-ontology model for integrating geographic information. Application of ontologies has also been explored in the field of geographic information retrieval \cite{r50}. A different use appears in \cite{r26}, where an OWL extension has new primitives for modeling locations and spatial relationships with a geographic ontology. Extensions of existing ontologies have also appeared in this knowledge domain. Thus in \cite{r49}, the authors propose a geographic ontology based on GML (Geography Markup Language) \cite{r41} and extends the OWL-S profile to geographic profiles. Another case is an extension of the NASA SWEET (Semantic Web for Earth and Environmental Terminology) ontology, which includes part of the hydrogeology domain \cite{r52}. 

As observed, an object-oriented approach has been used for modeling the different ontologies, selecting two representations, one visual with UML and the other textual with OWL. In the literature, there are many studies of environmental modeling which also make use of this approach (some of them are summarized in Table \ref{t13}). Some researchers suggest the use of UML as a language for geographical representation of ontological modeling \cite{r14, r23, r33}, showing it to be a resource for information exchange among specialists that is easy to read and interpret. In some cases, the UML models need to be transformed into ontological representations using specific model transformation tools. In \cite{r24}, the author defines XSLT transformations to construct an ontology in XMI format based on a logic model in UML. Another example of model transformation is OUP (Ontology UML Profile) \cite{r20}, which enables ODM/MDA standards to be used. XSLT is also used to transform OUP ontologies into OWL.  In \cite{r25}, the authors generate GML models from conceptual UML models which comply with the ISO 19118 standard. ONTOMET \cite{r10} is a UML-based framework for geospatial information communities which provides a flexible environment and enable interoperation of formal metadata specifications, domain vocabularies and extensions. GSIP (GEOINT Structure Implementation Profile) and GEOUML are two UML modeling approaches based on the ISO/TC211 standard: GSIP \cite{r42} uses a UML data model based on ISO/TC211 data models for characteristics, spatial geometry, topology, time and cover, and GEOUML \cite{r9} is a UML framework for conceptual modeling of GIS based on ISO 19100* standards. 

On the other hand, OWL \cite{r56} is a standard W3C language following an object-oriented philosophy for defining Web ontologies, although today it has spread widely and may be found in almost any domain. For its functioning, OWL relies on other languages such as XML, RDF, and RDF-S (RDF-Schema). There are also plenty of OWL-based approaches in the literature for defining ontological models in EMIS. In particular, HYMPAS-OWL \cite{r1} and ETHAN \cite{r47} should be mentioned because they are applied to ecological domains (more related to the SOLERES system domain). The first consists of three different applications for hydrological modeling. It facilitates layers with spatial information on a map which includes groundwater, roads, railroads, borders, streams and lakes, and covers all of the United States except Alaska. The ETHAN ontologies concentrate on biological taxonomies and characteristics associated with natural history. 

\begin{table}[!b]
\centering
\caption{UML and OWL approaches for environmental modeling}
\label{t13}
 {\fontsize{6.5pt}{8pt}\selectfont
\begin{tabular}{p{1.9cm}p{1.5cm}p{2.3cm}p{3.1cm}p{2.8cm}}
\hline
Name & OO Approach & Technologies & Organization & Domain\\
\hline
Gasevic &  UML & XSLT, Model Transform. & Belgrade Univ. & General Ontology\\
Pan &  UML & Model Transform. & IBM China Research Lab& General Ontology\\
OUP & UML and OWL & ODM, MDA, XSLT & Belgrade Univ. & General Ontology\\
Gronmo et al. &  UML & GML, ISO 19188 & SINTEF Telecom \& Informatics & Geographical Info.\\
ONTOMET &  UML and OWL & ISO 19115 & Univ. Drexel, USA & Geospatial \& Hydrology Info. \\
GSIP &  UML and OWL & ISO/TC211 & Nat. Geospatial-Intell. Agency & Geospatio-temporal Info.\\
GEOUML &  UML & ISO/TC211, ISO 19100* & Dpt. Informatics, U.Verona, Italia & GIS\\
Geographical-Space &  UML & Semantic Reasoning & Univ. Wuhan, China & Geospatial Info. \\
GeoOWL &  OWL & ISO/TC211 & W3C & Geographical Info.\\
HYMAPS-OWL &  OWL & Decision support system & Univ. Purdue, USA & Ecological \& Hydrolog. Info. \\
\hline
\end{tabular}
}
\end{table}

As expected, the progress of technologies has positively influenced the design and development of WIS.  Technologies such as trading services, ontologies and software agents have been incorporated in them offering a wide range of possibilities. Since the domain of the case study analyzed in this article is the environment, a rather exhaustive review was made of systems in this field which have been designed based on them. They can all be classified in one of three categories: environmental information management, assistance in decision-making for solving environmental problems, and simulation. Table \ref{t14} shows a comparison of systems related to the SOLERES system studied. Characteristics kept in mind were: (a) use of trading mechanisms, (b) use of modeling and design of ontologies, (c) incorporation of some type of interface agent or similar, and (d) the technology employed for its implementation. This table shows that none of the systems except SOLERES and BUSTER \cite{r55} use a trading service as such, and even in BUSTER this trading service is only partially implemented. And except for InfoSleuth \cite{r27}, where the agents use ontologies to describe the services they offer and the information to which they have access, they do not use ontologies like SOLERES either. In EDEN-IW \cite{r21}, the ontologies were modeled following Semantic Web technology rules, and again, BUSTER, which uses a hybrid ontology where each information source has an associated ontology to define its semantics. With regard to the use of agents, all the systems except FSEP \cite{r18} and BUSTER, have some type of interface agent that mediates for the user. It should be mentioned that SOLERES is the only system that combines all three of these characteristics. Finally, the use of multiple and varied technologies for the implementation of these systems, such as Java/C++, JADE, XML/RDF, DAML+OIL, etc., should be mentioned. 

\begin{table}
\centering
\caption{Comparison of EMIS architectures}
\label{t14}
 {\fontsize{7pt}{8pt}\selectfont
\begin{tabular}{llllll}
\hline
System & Trading & Ontologies & User Agent & Technologies & Domain\\
\hline
InfoSleuth & No & Yes & Yes & XML/RDF, KQML, OKBC & Water resources\\
EDEN-IW & No & Yes & Yes & JADE, DAML+OIL & Water resources\\
NZDIS & No & No & Yes & CORBA/OQL, MOF & Environmental data\\
FSEP & No & No & No & JACK & Metheorology\\
MAGIC  & No & No & Yes & FIPA-ACL, CORBA & Water management\\
DIAMON & No & No & Yes & Java/C++, FIPA-ACL & Water management\\
BUSTER & Yes & Yes & No & OIL, FIPA-OS & Geographic information\\
\hline
SOLERES & Yes & Yes & Yes & JADE, OWL, SPARQL, UML & Ecology\\
\hline
\end{tabular}
}
\end{table}

\section{Conclusions and future work}
\label{sec:conclus}

The exponential growth of information and information sources in large Web-based Information Systems hinders management tasks and information processing. Information search and retrieval operations are especially costly when working with heterogeneous database which because of their distributed nature have grown over time. In this context, the recent Big Data concept attempts to solve problems of data management and analysis in systems managing a large and varied volume of information and information sources. One example of this occurs in many current geographic and environmental information management systems, where large amounts of data are generated and have to be managed for studies and simulations for decision-making. Over time, these systems end up supporting large volumes of environmental data which make information search and retrieval operations very costly. 

This article presented an ontological trading model called OntoTrader, which evolved from the traditional ODP (\textit{Open Distributed Processing}) model, extending it to management not only of services, but of any type of information, which uses a ``Query-Search/Recovery-Response'' mechanism to improve information search and retrieval in Big Data systems. An architecture with two important innovations is proposed for its design. The first is the use of a document repository which could be instances of any ontology representing the metadata of the information in a specific domain, and the second is definition of an associated service/process ontology for each of the trading service interfaces, which improves the communication model and interaction with it. 

This article reported on modeling and formalizing of an ontological trading framework called TKRS (\textit{Trading-based Knowledge Representation System}) for Web-based Information Management Systems that handle large volumes of data. This framework was constructed following MDE (\textit{Model-Driven Engineering}) and ODE (\textit{Ontology-Driven Engineering}) guidelines and rules, which enable the architectural definition of the system to be separated from its implementation. The TKRS considers the system a distributed architecture comprised of a set of functional nodes (or ambients). A graphical editor developed in GMF (\textit{Graphical Modeling Framework}) makes it possible to generate the architecture of the system model to be deployed during the design stage. System node functionality is defined in the design stage as a type of capability the system must have, and can be implemented in different programming languages in the development stage. A system component implementation repository model and a system configuration model which connects each component in the architecture to its corresponding implementation were defined in the TKRS framework. TKRS uses M2M (\textit{model-to-model}) transformations to obtain a system configuration model that relates architecture models to implementations. Based on the configuration model, a concrete instance of the system (executable code) was obtained for the specific platform selected by applying an M2T (\textit{model-to-text}) transformation.

Finally, the article presented a real case study, the design and implementation of SOLERES-KRS, a subsystem responsible for managing knowledge in the SOLERES environmental management system, which handles large volumes of cartographic maps, satellite images and ecological maps. The TKRS guidelines were followed for its construction and the software agent technology was selected for its implementation. Due to the nature of the TKRS framework, two data ontologies were used, the EIM ontology to represent the environmental information metadata, and the EID ontology to represent the meta-metadata used by the trading agents that implement the OntoTrader model. Communication and coordination of the different agents make use of three service/process ontologies. The system was designed with the TKRS model editor developed in GMF and a JADE implementation of the Multi-Agent System. The ontologies used by the agents for communicating and interacting, and the ontologies representing the system knowledge are expressed in OWL/XML, previously designed using UML modeling techniques. 

As future lines of work, in the first place, it would be of interest to modify the \textit{stand-alone} design of the OntoTrader trading service to implement the Link and Proxy interfaces, along with the corresponding service/process ontologies, so it would be \textit{full-service} trader. In the second place, it is intended to use the trading service to automatically establish the connection between architectural elements and their corresponding implementations. At present, this is being supervised by the DSL defined. This would imply that the trading service itself would independently locates the best components for TKRS implementation automatically, based on the definition of the system architecture with the graphical editor. Finally, it may be useful to have an Integrated Development Environment (IDE) for design of implementation repository models and an application that would integrate all the tools used in the TKRS framework design and implementation tasks.

\section*{Acknowledgements}

\noindent This work was funded by the EU ERDF and the Spanish Ministry of Economy and Competitiveness (MINECO) under Project TIN2013-41576-R and the Andalusian Regional Government (Spain) under Project P10-TIC-6114.


\begin{thebibliography}{99}


\bibitem{r1}
ABED (2010). A web GIS for hydrolic model operation. Agricultural \& Biological Engineering Department, Purdue University. https://engineering.purdue.edu/mapserve/LTHIA7/index.html.

\bibitem{r2}
Albert, M., L\"{a}ngle, T., Woern, H., and FR, K.U.G. (2002). Development tool for distributed monitoring and diagnosis systems. Defense Technical Information Center.

\bibitem{alexander}
Alexander, F.J., Adolfy, H., and Szalay, A. (2011). special issue on Big Data. Computing in Science \& Engineering, 13(6).

\bibitem{r3}
An, Y. and Zhao, B. (2007). Geo ontology design and comparison in geographic information integration. In Fourth Int. Conference on Fuzzy Systems and Knowledge Discovery (FSKD), volume 4, pages 608--612. IEEE Computer Society. 

\bibitem{r4}
Asensio, J. A., Iribarne, L., Padilla, N., and Ayala, R. (2008). Implementing trading agents for adaptable and evolutive UI-COTS components architectures. In Proceedings of the International Conference on e-Business (ICE-B), pages 259--262, Porto, Portugal. INSTICC Press. 

\bibitem{r5}
Asensio, J. A., Iribarne, L., Padilla, N., and Vicente-Chicote, C. (2011). A model-driven approach for deploying trading-based knowledge representation systems. In On the Move to Meaningful Internet Systems: OTM 2011 Workshops, volume 7046 of Lecture Notes in Computer Science, pages 180--189. Springer.  

\bibitem{r6}
Asensio, J. A., Padilla, N., and Iribarne, L. (2012). An ontology-driven case study for the knowledge representation of management information systems. In Information Systems, E-learning, and Knowledge Management Research, volume 278 of Communications in Computer and Information Science (CCIS), pages 426--432. Springer.  

\bibitem{bakshi2012} 
Bakshi, K. (2012). Considerations for big data: Architecture and approach. In IEEE Aerospace Conference, pages 1--7. IEEE.

\bibitem{bayard}
Bayard Cushing, J. (2013). Beyond Big Data? Computing in Science \& Engineering, 15(5):4--5.

\bibitem{r7}
Bearman, M. (1997). Tutorial on ODP trading function. Faculty of Information Sciences Engineering. University of Canberra. 

\bibitem{r8}
Beitz, A. and Bearman, M. (1994). An ODP trading service for DCE. In Proceedings of the First International Workshop on Services in Distributed and Networked Environments, pages 42--49. IEEE. 

\bibitem{r9}
Belussi, A., Negri, M., and Pelagatti, G. (2004). GeoUML: A geographic conceptual model defined through specialization of ISO TC211 standards. In Proceedings of the 10th EC GI \& GIS Workshop, pages 1--10, Warsaw, Poland. 

\bibitem{r10}
Bermudez, L. E. (2004). Ontomet: Ontology metadata framework. Philadelphia, PA. Drexel University. 

\bibitem{r11}
Brilhante, V. B. (2004). An ontology for quantities in ecology. In Advances in Artificial Intelligence - 17th Brazilian Symposium on Artificial Intelligence (SBIA), Proceedings, Lecture Notes in Computer Science, pages 144--153. Springer. 

\bibitem{r12}
Ceccaroni, L., Cort\'es, U., and S\`anchez-Marr\`e, M. (2004). OntoWEDSS: Augmenting environmental decision-support systems with ontologies. Environmental Modelling and Software, 19(9):785--797. 

\bibitem{r13}
Collins, J., Ketter, W., and Gini, M. L. (2009). Flexible decision control in an autonomous trading agent. Electronic Commerce Research and Applications, 8(2):91--105. 

\bibitem{r14}
Cranefield, S. and Purvis, M. (1999). UML as an ontology modelling language. In Proc. of the Workshop on Intelligent Information Integration, 16th International Joint Conference on Artificial Intelligence (IJCAI), pages 46--53. 

\bibitem{r15}
Craske, G., Tari, Z., and Kumar, K. R. (1999). DOK-Trader: A CORBA persistent trader with query routing facilities. In Distributed Objects and Applications (DOA), pages 230--240. 

\bibitem{r16}
Criado, J., Padilla, N., Iribarne, L., and Asensio, J. A. (2010). User interface composition with COTS-UI and trading approaches: Application for Web-based Environmental Information Systems. In Knowledge Management, Information Systems, E-Learning, and Sustainability Research, Part I, volume 111 of Communications in Computer and Information Science (CCIS), pages 259--266. Springer.  

\bibitem{r17}
Criado, J., Padilla, N., Iribarne, L., Asensio, J. A., and Mu\~noz, F. (2010). Ontological trading in a multi-agent system. In Trends in Practical Applications of Agents and Multiagent Systems - 8th International Conference on Practical Applications of Agents and Multiagent Systems (PAAMS), Special Sessions and Workshops, volume 71 of Advances in Intelligent and Soft Computing (AISC), pages 321--329, Salamanca, Spain. Springer.  

\bibitem{r18}
Dance, S., Gorman, M., Padgham, L., and Winikoff, M. (2003). An evolving multi agent system for meteorological alerts. In: Proceedings of the second international joint conference on Autonomous agents and multiagent systems, pp. 966--967. ACM New York, NY, USA.

\bibitem{r19}
Di Lecce, V., Pasquale, C., and Piuri, V. (2004). A basic ontology for multi agent system communication in an environmental monitoring system. In IEEE International Conference on Computational Intelligence for Measurement Systems and Applications (CIMSA), pages 45--50, Boston, MA. Institute of Electrical and Electronics Engineers. 

\bibitem{r20}
Djuric, D., Gasevic, D., Devedzic, V., and Damjanovic, V. (2005). A UML profile for OWL ontologies. In Model Driven Architecture, European MDA Workshops: Foundations and Applications (MDAFA), volume 3599 of Lecture Notes in Computer Science, pages 204--219. Springer. 

\bibitem{r21}
EDEN-IW (2001). Environmental data exchange network for inland water. Tech. rep., http://www.eden-iw.org.

\bibitem{r22}
EMF (2013). Eclipse Modeling Framework. http://www.eclipse.org/modeling/emf/. 

\bibitem{r23}
Falkovych, K., Sabou, M., and Stuckenschmidt, H. (2003). UML for the semantic web: Transformation-based approaches. In Knowledge Transformation for the Semantic Web, pages 92--106. 

\bibitem{r24}
Gasevic, D., Djuric, D., Devedzic, V., and Damjanovic, V. (2004). Converting UML to OWL ontologies. In Proceedings of the 13th International Conference on World Wide Web - WWW (Alternate Track Papers \& Posters), pages 488--489. ACM. 

\bibitem{r25}
Gronmo, R., Solheim, I., and Skogan, D. (2002). Experiences of UML-to-GML encoding. In Proceedings of the 5th AGILE Conference, pages 1--8. 

\bibitem{r26}
Huang, M. (2008). A new method to formal description of spatial ontology. Information Technology and Environmental System Sciences, (3):417--421. 

\bibitem{r27}
InfoSleuth (2005). The infosleuth agent system. Tech. rep., http://www.argreenhouse.com/InfoSleuth/.

\bibitem{r28}
Iribarne, L., Asensio, J. A., Padilla, N., and Ayala, R. (2008). SOLERES-HCI: Modelling a Human-Computer Interaction framework for open EMS. In The Open Knowledge Society, volume 19 of Communications in Computer and Inf. Science (CCIS), pages 320--327. Springer.  

\bibitem{r29}
Iribarne, L., Padilla, N., Asensio, J. A., Mu\~noz, F., and Criado, J. (2010). Involving web-trading agents \& MAS -- an implementation for searching and recovering environmental information. In Int. Conference on Agents and Artificial Intelligence (ICAART), volume 2, pages 268--273, Valencia, Spain. INSTICC Press. 

\bibitem{r30}
Iribarne, L., Padilla, N., Criado, J., Asensio, J. A., and Ayala, R. (2010). A model transformation approach for automatic composition of COTS user interfaces in web-based information systems. Journal of Information Systems Management (JISM), 27(3):207--216.  

\bibitem{r31}
Iribarne, L., Criado, J., Padilla, N., and Asensio, J. A. (2011). Using COTS-widgets architectures for describing user interfaces of Web-based Information Systems. International Journal of Knowledge Society Research (IJKSR), 2(3):61--72.  

\bibitem{r31bis}
Iribarne, L., Padilla, N., Ayala, R., Asensio, J.A., Criado, J. (2014). OntoTrader: An Ontological Web-Trading agent approach for Environmental Information retrieval. The Scientific World Journal, Volume 2014 (2014), Article ID 560296, 25 pages.

\bibitem{ishwarappa2015brief}
Ishwarappa, Anuradha, J. (2015). A Brief Introduction on Big Data 5Vs Characteristics and Hadoop Technology. Procedia Computer Science, 48:319-324.

\bibitem{r32}
ITU-T and ISO/IEC (1998). Information Technology -- Open Distributed Processing -- ODP Trading Function -- Part 1: Specification. ITU-T Rec. X.950 -- ISO/IEC 13235-1. 

\bibitem{kaisler2014introduction}
Kaisler, S., Armour, F., Espinosa, J.A. (2014). Introduction to big data: Challenges, opportunities, and realities minitrack. In 47th Hawaii International Conference on System Sciences (HICSS), pages 728. IEEE.

\bibitem{r33}
Kogut, P., Cranefield, S., Hart, L., Dutra, M., Baclawski, K., Kokar, M., and Smith, J. (2002). UML for ontology development. Knowledge Engineering Review, 17(1):61--64. 

\bibitem{kune2016anatomy}
Kune, R., Konugurthi, P.K., Agarwal, A., Chillarige, R.R., Buyya, R. (2016). The anatomy of big data computing. Software: Practice and Experience, 46(1):79--105.

\bibitem{r34}
Kutvonen, L. (1996). Overview of the DRYAD trading system implementation. In The IFIP/IEEE International Conference on Distributed Platforms: Client/Server and Beyond: DCE, CORBA, ODP and Advanced Distributed Applications - ICDP, pages 314--326. Chapman \& Hall. 

\bibitem{r35}
Lamparter, S. and Schnizler, B. (2006). Trading services in ontology-driven markets. In ACM Symposium on Applied Computing (SAC), pages 1679--1683. 

\bibitem{r36}
Lieberman, J., Singh, R., and Goad, C. (2007). W3C Geospatial Ontologies. W3C Incubator Group Report. http://www.w3.org/2005/Incubator/geo/XGR-geo-ont-20071023/. 

\bibitem{madden2012from}
Madden, S. (2012). From databases to big data. IEEE Internet Computing, 16(3):4--6.

\bibitem{r37}
Mathe, J. L., Werner, J., Lee, Y., Malin, B., and Ledeczi, A. (2008). Model-based design of clinical information systems. Methods of Information in Medicine, 47(5):399--408. 

\bibitem{r38}
Merz, M., M\"{u}ller, K., and Lamersdorf, W. (1994). Service trading and mediation in distributed computing systems. In Int. Conf. on Distributed Computing Systems (ICDCS), pages 450--457. 

\bibitem{r39}
M\"{u}ller, K., Merz, M., and Lamersdorf, W. (1995). The TRADEr: Integrating trading into DCE. In The 3rd IFIP Conference on Open Distributed Processing, pages 476--487. Chapman \& Hall. 

\bibitem{r40}
NZDIS (2001). New Zealand Distributed Information Systems project. Tech. rep., http://nzdis.otago.ac.nz.

\bibitem{r41}
OGC (2007a). Geography Markup Language (GML). Open Geospatial Consortium. http://www.opengeospatial.org/standards/gml/. 

\bibitem{r42}
OGC (2007b). GEOINT structure implementation profile schema processing. Open Geospatial Consortium. 

\bibitem{r43}
Okawa, T., Kaminishi, T., Hirabayashi, S., Koizumi, H., and Sawamoto, J. (2008). An information system development method based on the link of business process modeling with executable UML modeling and its evaluation by prototyping. In AINA Workshops, pages 1057--1064. 

\bibitem{r44}
OOC (2001). ORBacus trader. ORBacus for C++ and Java. 

\bibitem{r45}
Pan, Y., Xie, G. T., Ma, L., Yang, Y., Qiu, Z., and Lee, J. (2006). Model-Driven Ontology Engineering. In J. Data Semantics VII, volume 4244 of Lecture Notes in Computer Science, pages 57--78. Springer. 

\bibitem{r46}
PrismTech (2001). OpenFusion trading service whitepaper. PrismTech. http://www.prismtech.com/openfusion/re\-sources/white-papers/. 

\bibitem{r47}
Project Spire (2003). ETHAN. http://www.csee.umbc.edu/csee/research/spire/. 

\bibitem{quilitz2008sparql}
Quilitz, B., and Leser, U. (2008). Querying distributed RDF data sources with SPARQL. In The Semantic Web: Research and Applications, volume 5021 of Lecture Notes in Computer Science, pages 524--538. Springer Berlin Heidelberg.


\bibitem{r48}
Schmidt, D. C. (2001). The ADAPTATIVE communication environment (TAO): AceORB. University of California. http://www.theaceorb.com/. 



\bibitem{r49}
Shen, J., Krishna, A., Yuan, S., Cai, K., and Qin, Y. (2008). A pragmatic gis-oriented ontology for location based services. In Australian Software Engineering Conference (ASWEC), pages 562--569. IEEE Computer Society. 

\bibitem{sheth2014deriving}
Sheth, A. (2014). Transforming Big Data into Smart Data: Deriving value via harnessing Volume, Variety, and Velocity using semantic techniques and technologies. In IEEE 30th International Conference on Data Engineering (ICDE), pages 2. IEEE.

\bibitem{r50}
Song, J., Zhu, Y., and Wang, J. (2007). A study of semantic retrieval system based on geo-ontology with spatio-temporal characteristic. In Proceedings of International Symposium on Distributed Computing and Applications to Business, Engineering and Science (DCABES), volume I-II, pages 1029--1034. 

\bibitem{r51}
Taniar, D. and Rahayu, J. W. (2004). Web Information Systems. Print. IGI Global, Hershey. 

\bibitem{r52}
Tripathi, A. and Babaie, H. A. (2008). Developing a modular hydrogeology ontology by extending the SWEET upper-level ontologies. Computers \& Geosciences, 34(9):1022--1033. 

\bibitem{r53}
Tsai, W.-T., Huang, Q., Xu, J., Chen, Y., and Paul, R. A. (2007). Ontology-based dynamic process collaboration in service-oriented architecture. In Service-Oriented Comp. and App. (SOCA), pages 39--46. 

\bibitem{r54}
Vasudevan, V. and Bannon, T. (1999). WebTrader: Discovery and programmed access to web-based services. In Poster at the 8th International WWW Conference (WWW8), Toronto, Canada. 

\bibitem{r55}
V\"{o}gele, T., H\"{u}bner, S., and Schuster, G. (2003). BUSTER ? An Information Broker for the Semantic Web. KI-K\"{u}nstliche Intelligenz 3(3), 31--34.

\bibitem{r56}
W3C (2004a). OWL Web Ontology Language reference. World Wide Web Consortium. http://www.w3.org/OWL/. 

\bibitem{r57}
Wang, Y., Dai, J., Sheng, J., Zhou, K., and Gong, J. (2007). Geoontology design and its logic reasoning. In SPIE Proceedings - Geospatial Information Science, volume 6753, pages 675309-1 -- 675309-11. 

\bibitem{r58}
W\"{o}rn, H., L\"{a}ngle, T., and Albert, M. (2002). Multi-agent architecture for monitoring and diagnosing complex systems. In: The 4th International Workshop on Computer Science and Information Technologies (CSIT) University of Patras/Greece.

\bibitem{r59}
Xiao-feng, M., Bao-wen, X., Qing, L., Ge, Y., Jun-yi, S., Zhengding, L., and Yan-xiang, H. (2006). A survey of web information technology and application. Wuhan University Journal of Natural Sciences, 11:1--5. 

\bibitem{r60}
Zhan, Q., Li, D., and Shao, Z. (2007). An architecture for ontology-based geographic information semantic grid service. In Geoinformatics 2007: Geospatial Information Technology and Applications, SPIE Proceedings, volume 6754, pages 67541U -- 67541U-9. 

\bibitem{r61}
Ziming, Z. and Liyi, Z. (2007). An integrated approach for developing e-commerce system. In International Conference on Wireless Communications, Networking and Mobile Computing (WiCom), pages 3596--3599.


\end{thebibliography}
\end{document}